\documentclass[prc,nofootinbib,showpacs]{revtex4}

\usepackage{graphicx}
\usepackage[usenames,dvipsnames]{color}
\usepackage{amsmath,amssymb,bbold}
\usepackage{epsfig,rotate,latexsym}
\usepackage{subfig}
\usepackage{hyperref}
\usepackage{comment}

\begin{document}

\title{In medium properties of axion within a Polyakov loop enhanced Nambu-Jona-Lasinio model}

\author{Arpan Das$^{1}$}
\email{arpan.das@ifj.edu.pl}
\author{Hiranmaya Mishra$^{2}$}
\email{hm@prl.res.in}
\author{Ranjita K. Mohapatra${^3}$}
\email{ranjita.iop@gmail.com}

\affiliation{$^{1}$Institute of Nuclear Physics Polish Academy of Sciences, PL-31-342 
Krakow, Poland}
\affiliation{$^{2}$Theory Division, Physical Research Laboratory, 
Navrangpura, Ahmedabad 380 009, India}
\affiliation{$^{3}$ Department of Physics, Banki Autonomous College, Cuttack 754008, India.}
 \date{\today}

\begin{abstract}
We estimate the axion properties i.e. its mass, topological susceptibility and the self-coupling within 
the framework of Polyakov loop enhanced Nambu-Jona-Lasinio (PNJL) model at finite temperature and quark 
chemical potential. PNJL model, where quarks couple simultaneously to the chiral condensate and to a background 
temporal quantum chromodynamics (QCD) gauge field, includes two important features of QCD phase transition,
 i.e. deconfinement 
and chiral symmetry restoration. The Polyakov loop in PNJL model 
plays an important role near the critical temperature. We have shown significant difference in the 
axion properties calculated in PNJL model compared to  the same obtained using
 Nambu-Jona-Lasinio (NJL) model. We find that 
both the mass of the axion  and its self-coupling are correlated with the chiral transition as well as
the confinement-deconfinement transition. We have also estimated the axion properties at finite chemical potential.
Across the QCD transition temperature and/or quark chemical potential 
axion mass and its self-coupling also changes significantly. Since the PNJL model includes both the 
fermionic sector and the gauge fields, it can give reliable estimates of the axion properties, 
i.e. it's mass and the self-coupling in a hot and dense QCD medium. We also compare our results with the lattice QCD results whenever available.
\end{abstract}

\maketitle

\section{INTRODUCTION}
\label{intro}
The axion was originally predicted to solve the strong CP (charge conjugation and parity) problem in a dynamical way \cite{Peccei1977prl,Peccei1977prd,Weinberg1978prl, Wilczek1978prl}. Due to the non abelian nature of the gauge fields QCD allows topologically non-trivial Chern-Simons term, $\mathcal{L}_{\theta}\sim \theta_{}~\text{Tr} G_{\mu\nu}\tilde{G}^{\mu\nu}$. Note that this Chern-Simons term which is allowed by the gauge symmetry, does not affect the classical equation of motion. However, this term has important quantum mechanical effects \cite{Coleman1979}. For a nonvanishing value of $\theta$, CP symmetry is explicitly broken in QCD. Stringent constraints 
on the CP-violating $\theta$ term comes from the measurement of the electric dipole moment (EDM) of neutron, i.e. $\theta\lesssim 10^{-11}$ \cite{Crewther1979, Pendlebury}. Note that weak interaction is CP violating and the physically measurable $\theta$ parameter has two contributions coming from QCD as well as weak interaction. So a naive argument of CP symmetry is not a good explanation of the smallness of the physical $\theta$ parameter. The smallness of $\theta$ implied by the EDM constraint is a fine-tuning problem involving a precise cancellation between two dimensionless terms generated by physics at different scales. Spontaneous breaking of the Peccei-Quinn (PQ) symmetry is an elegant and robust manner to solve the strong CP problem in a dynamical manner which predicts the smallness of the $\theta$ \cite{Peccei1977prl,Peccei1977prd}. Spontaneous breaking of Peccei-Quinn (PQ) symmetry also naturally predicts a pseudo-Goldstone boson which is known as the axion \cite{Cortona,KimCarosi}. In the original axion model formulated by Peccei and Quinn\cite{Peccei1977prl,Peccei1977prd}, Weinberg \cite{Weinberg1978prl} and Wilczek \cite{Wilczek1978prl}, the spontaneous breaking of the $U(1)_{PQ}$ symmetry occurs simultaneously with the electroweak symmetry breaking giving rise to observational signals of axions which is in contradiction with the observational evidence, e.g. $K,J/\psi$ meson decay \cite{Kim1987}. These constraints can be avoided in the invisible axion models where the PQ symmetry breaking scale is high which gives rise to very light and weakly interacting axion. \cite{Kimprl,Shifman1980}.  

Axion mass, coupling to the other particles are inversely proportional to the PQ symmetry breaking scale. Hence axions are very light as well as weakly interacting particles and it has been considered as a candidate for dark matter \cite{Duffy2009,Turner1991,Visinelli2009,raffelt1990,axionsun,cheng1988,kim1987,marsh2016}. The effect of axion on the stellar evolution has also been considered in the literature. Due to the 
strangeness-changing non-leptonic weak interaction during the conversion of a neutron star into a strange star, axions could be copiously produced. Axion might drastically alter the energy budget of stars \cite{SuhLee}. Axions can be produced in hot and dense astrophysical plasma which can transport energy out of stars. One can put a strong constraint on the axion properties i.e. mass, coupling with normal matter and radiation using the stellar-evolution lifetimes or energy-loss rates observation \cite{Raffelt1991,Janka1996,Iwamoto1984,Umeda1998}.
For stellar longevity, the rate at which a star can liberate its nuclear-free energy is important, which is not only controlled by the nuclear reaction rates but also depends on the rate at which nuclear energy can be transported through the star and radiated into the vacuum. In the absence of weakly interacting low mass (as compared to typical stellar temperature) particles, by photon cooling the energy is dissipated away from stars. However, a
weakly interacting low mass particle has the potential to efficiently transporting energy away and thereby shortening the lifetime of stars. Note that for efficient transport of energy liberated in the nuclear reactions in a star, the 
particle should be weakly interacting, but the interaction with the nuclear matter should not be very small so that the weakly interacting particles should be
produced in sufficient numbers to carry away the nuclear energy efficiently. For stars hotter than about $10^8$ Kelvin neutrino cooling becomes comparable with the photon cooling. Being a small mass and weakly interacting in nature, axions if produced in the hot and dense medium inside the stars can also acts as a coolant. In stars the axions are produced by, Compton like process $(\gamma + e^-\rightarrow a+ e^-)$,  the Primakoff process $(\gamma + Z(e^-)\rightarrow a+Z(e^-))$ and axion bremsstrahlung $(e^-+Z\rightarrow a+e^-+Z)$. The axion emissivity due to these processes is proportional to the axion mass ($m_a^2$) \cite{axionsun}. Therefore estimation of axion mass is of great importance to investigate the effect of axion on the stellar cooling.
Further it has been suggested that axion can form stars as well as a Bose-Einstein condensate \cite{axionbec1,axionbec2,axionbec3,axionbec4,axionbec5,axionbec6,axionbec7,axionbec8,axionbec9,axionbec10,axionbec11,axionbec12,axionbec13}. In the context of 
Bose-Einstein condensation of dark matter axions, to estimate the thermalization of axion, it's self interaction plays an important role in the calculation of the relaxation rate \cite{axiorelaxation}. Further ultra-light axion self-interactions
can play an important role on the large scale structure of the Universe \cite{riotto}. 
It is therefore of paramount importance to 
know the
characteristics
 of axion properties, e.g. axion mass and the self-coupling in a hot and dense medium. 
 
It is important to note that 
there  are  various  computations  of finite temperature axion mass which are available in literature, e.g. dilute instanton gas, lattice QCD, 
instanton liquid model etc, which may not agree with one another
 \cite{axionmass1,axionmass2,axionmass3,axionmass4,axionmass5,axionmass6}. The temperature dependence
  of axion mass is important which can affect significantly the axion dark matter abundance. Estimation of the 
axion potential closer to the QCD transition scale is also important and in the absence of this knowledge, 
one parameterize the axion mass in a way that resembles the result from the dilute instanton gas model
 \cite{dilute1,dilute2}. It may also be noted that such a parameterization of the axion mass with 
temperature can have a discontinuity that does not encapsulate the variation of axion mass across the 
QCD transition scale \cite{dilute3}. At relatively high
temperatures with respect to the quark hadron transition scale, one can use perturbative techniques to 
estimate axion properties, e.g., the dilute instanton gas approximation, however around and below the QCD 
transition scale, nonperturbative effects can have a significant impact on the axion mass and coupling. To 
estimate the response of the axion to a QCD thermal medium
one can use QCD inspired effective field theories and phenomenological models, e.g. 
the chiral perturbation theory ($\chi$PT) \cite{chiptrev1,chiptrev2,chiptrev3}, Nambu-Jona-Lasinio (NJL) model 
\cite{BuballaReview,KlevanskyReview}, etc. $\chi$PT which has been used to study the $\theta$ vacuum of QCD and 
QCD axion physics, can predict a value of topological susceptibility (topological susceptibility is proportional 
to the axion mass) which agrees with the lattice QCD results at zero temperature 
\cite{chipt1,chipt2,chipt3,chipt4,chipt5,chipt6,chipt7}.
 Further, the computation of the axion potential can be extended to finite temperature for chiral Lagrangian. 
In particular, at temperatures below the QCD transition scale ($\sim 170$ MeV) using chiral Lagrangian the temperature 
dependence of the axion potential and its mass can be estimated. Although the $\chi$PT can give rise to 
reliable prediction at low temperature, at high temperature $\chi$PT results may not be reliable, because in 
$\chi$PT there are no partonic degrees of freedom which becomes important near and above the QCD transition temperature. 
Perturbative expansion is also not under control around the QCD transition scale and non-perturbative methods 
are required to study the axion properties.

Because of the limitation of $\chi$PT and perturbative techniques one can use QCD inspired chiral effective models, e.g. Nambu-Jona-Lasinio (NJL) model to investigate the thermal properties of the axion. NJL model has been used earlier to study the CP-violating effects and effect of theta vacuum on the QCD phase diagram \cite{NJLcpviolation1, NJLcpviolation2,NJLcpviolation3,NJLcpviolation4,NJLcpviolation5}. 
This approach has been considered in Refs.\cite{NJLaxion1} and Ref.\cite{NJLaxion2} to study the axion mass and self-coupling at finite temperature in the absence
as well as in the presence of a magnetic field. All these calculations show that near the chiral transition temperature the axion mass and self-coupling are significantly modified. 
The NJL model which effectively explains some of the key features of QCD e.g., chiral symmetry breaking and its restoration, does not address the effects of the gluon degrees of freedom adequately. In the NJL model the gluonic degrees of freedom are “integrated out” and they are replaced by a local four-Fermi type interaction of quark colour currents. Improvements upon the NJL model e.g., the
Polyakov loop enhanced Nambu-Jona-Lasinio (PNJL) model takes into account this missing feature by including
a temporal background gluon field. As a result, both chiral and deconfinement aspects of QCD are captured within a single framework \cite{rattiPNJL1,rattiPNJL2,rattiPNJL3,rattiPNJL4,fukushimaPNJL1,sanjayghoshPNJL1,sanjayghoshPNJL2,sanjayghoshPNJL3}. In the PNJL model, the nonzero value of the Polyakov loop around the QCD transition scale plays an important role, which may be important to study the axion physics, particularly across the QCD transition scale.  

Therefore in the present article, we 
study the axion mass and self-coupling using the Polyakov loop enhanced Nambu-Jona-Lasinio (PNJL) model. The two most important properties of QCD are the chiral transition and confinement-deconfinement transition. Hence for an effective description of the QCD 
medium near the transition temperature, the effective model should also reflect these properties of QCD. Nambu-Jona-Lasinio (NJL) model which is a key ingredient of the PNJL model only deals with the fermionic part without gauge fields. NJL model includes the global symmetries of QCD in the fermionic sector,e.g. chiral symmetry, baryon number, electric charge, strange number symmetries, etc. In the NJL model, the dynamical generation of mass due to the multi quark interactions leads to the spontaneous breaking of chiral symmetry. 
In the NJL model, the local $SU(3)_c$ gauge symmetry of  QCD  is replaced by a  
global $SU(3)_c$ symmetry. So the NJL model lacks confinement property due to the absence of the QCD gauge fields. Note that for $N_c = 2$ confinement is less significant and the thermodynamics can be described quite successfully using the simplest NJL model \cite{rattiPNJL5}. On the other hand for $N_c=3$ confinement is significant. In the PNJL model both the chiral condensate and the Polyakov loop are considered as classical homogeneous fields which couple to the quarks according to the symmetries and symmetry breaking patterns of QCD, therefore describes various  aspects of confinement and chiral symmetry breaking in a unifying framework.
The confinement-deconfinement transition which is characterized by the Polyakov loop order parameter is well defined in the static quark limit. 
Confinement-deconfinement transition is characterized by the spontaneous breaking 
of the $Z(3)$ center symmetry of QCD \cite{z3ref1,z3ref2,z3ref3,z3ref4}. However, in the presence of dynamical quarks, the center symmetry is explicitly broken. Hence in the presence of a dynamical quark Polyakov loop cannot be considered as an order parameter, but the Polyakov loop still serves as an indicator of the confinement deconfinement transition \cite{z3ref5,z3ref6,z3ref7,z3ref8,z3ref9,z3ref10}. 
It is important to note that PNJL model also has some limitations, e.g. in the PNJL model the Polyakov loop is considered as a simple static background field. Transverse gluons which play an important role in the thermodynamics of QCD matter at a very high temperature $T\gtrsim 2.5 T_c$, are not considered in the PNJL model. Therefore the PNJL model is expected to describe QCD thermodynamics only within a limited range of temperature \cite{transversegluon}.
The previous studies of axion properties within the framework of the NJL model indicate that QCD transition significantly modifies axion mass and self coupling\cite{NJLaxion1,NJLaxion2}. Axion mass and axion self-coupling decrease rapidly across the chiral transition temperature and there is a correlation between the quark-antiquark condensate and axion properties studied in this model. In the PNJL model at finite temperature presence of the nonvanishing value of the Polyakov loop affects quark-antiquark condensate, which eventually also affects the axion mass and its self-coupling at finite temperature.

This paper is organised in the following manner. After the introduction, in Sec.\eqref{formalism} we discuss the formalism to study the axion properties, i.e. it's mass and self-coupling in the PNJL model. Using the formalism as given in Sec.\eqref{formalism} we estimate the axion mass and self-coupling at finite temperature and quark chemical potential. We show the results and the discussions of these results in Sec.\eqref{results}.
Finally in Sec.\eqref{conclusion} we conclude our investigation with an outlook to it. 

\section{FORMALISM: AXION WITHIN THE PNJL MODEL}
\label{formalism}

QCD, in principle, can have a parity violating term the so called $\theta$ term 
\begin{equation}
\mathcal {L}_\theta=\frac{\theta g_s^2}{64\pi^2}G_{\mu\nu}^a\tilde G^{\mu\nu}_a
\end{equation}
where, $G_{\mu\nu}^a$ is the gluon field strength and $\tilde{G}_{\mu\nu}^a$ being its dual. Such term respects 
Lorentz invariance as well as gauge invariance but violates parity unless $\theta=0$mod $\pi$. However, in nature,
QCD respects CP to a large extent in vacuum as the the magnitude of the CP violating term $\theta$ is small $\theta < 0.7\times 10^{-11}$ arising from measurement of intrinsic electric dipole moment of neutrons. 

A dynamical and elegant way to explain the smallness of $\theta$ is to elevate 
$\theta$ to a field such a way that it has a vanishing vacuum expectation 
value and the normalized axion field is denoted as $a(x) =\theta(x)f_a $. Here $f_a$ is the axion decay constant which also represents the PQ symmetry breaking scale. Phenomenology of axion is controlled by the axion decay constant $f_a$. 
Astrophysical observations, e.g. cooling rate of the SN1987A supernova,black hole superradiance  put 
stringent bounds on the PQ symmetry breaking scale, $10^8 \lesssim f_a\lesssim 10^{17}$ GeV
\cite{fabound1,fabound2,fabound3,fabound4,fabound5,fabound6}. Typically one could consider $f_a$ to be of the order of grand unified scale (GUT scale) $\sim 10^{16}$ GeV. 
Hence the interaction between the axion field and the QCD gauge field now can be expressed as
$\mathcal{L}_{\theta}\propto (a/f_a)G\tilde{G}$. In vacuum for $\theta=0$, spontaneous parity violation does not exist
as per Vafa-Witten theorem \cite{wittenvafa}. On the other hand, there could be CP violation for $\theta=\pi$,
 by Dashen phenomena, with the appearance of two degenerate CP-violating vacua separated by a potential barrier. Because of the nonperturbative nature of this CP violating term, this has been studied in different low energy effective models. In the present investigation, we shall confine our attention to PNJL model.

The PNJL model which is an extension of NJL model, is defined by a Lagrangian which couples the quarks to a temporal background
gauge field representing Polyakov loop dynamics. Explicitly, the 
Lagrangian density of the two flavour PNJL model with the Kobayashi-Maskawa-’t Hooft determinant interaction 
term incorporating the interaction with the axion can be expressed as \cite{sasakiPNJL1,NJLcpviolation2},
\begin{align}
 \mathcal{L}_{PNJL} & = \bar{q}\left(i\gamma_{\nu}D^{\nu}-\hat{m}\right)q-\mathcal{U}(\Phi[A],\bar{\Phi}[A],T)+ g_1\sum_{a=0}^3\bigg[(\bar{q}\tau_a q)^2+(\bar{q}i\gamma_5\tau_a q)^2\bigg]\nonumber\\
 &~~~~~~~~~~~~~~~~~~~~ +8g_2\bigg[e^{ia/f_a}\text{det}(\bar{q}_Rq_L)+e^{-ia/f_a}\text{det}(\bar{q}_Lq_R)\bigg],
 \label{equ1}
\end{align}
here $q=(q_u,q_d)^T$ is the quark field, $\hat{m}$ represents the current quark mass 
matrix $\text{diag}(m_u,m_d)$. In the present investigation we consider $m_u=m_d=m$. $\tau_0$ is the $2\times 2$ identity matrix, $\tau_a$ with 
$a=1,2,3$ are the Pauli matrices. The covariant derivative 
$D^{\nu}=\partial^{\nu}-iA^{\nu}$ and $A^\nu=\delta^{\nu}_0A^0$. The gauge coupling is absorbed in the definition of $A_\mu=
g_s{\cal A}_\mu^a\frac{\lambda^a}{2}$, where ${\cal A}_\mu^a$ is the SU(3) gauge field and 
$\lambda_a$ is the Gell-Mann matrix, $g_s$ is the 
gauge coupling. 
In the NJL sector, $g_1$ 
denotes the coupling of the four-quark 
interaction which includes scalar and pseudoscalar type interactions. This interaction term 
is symmetric under $SU(2)_{L}\times SU(2)_{R}\times U(1)_V\times U(1)_A\times SU(3)_c$ 
symmetry. $g_2$ is the coupling of the Kobayashi-Maskawa -'t Hooft determinant interaction. 
This determinant is in the flavor space. The determinant interaction term explicitly breaks 
the $U(1)_A$ symmetry of the Lagrangian. The Polyakov loop potential 
$\mathcal{U}(\Phi,\bar{\Phi},T)$, is the effective potential of the traced Polyakov 
loop and its Hermitian conjugate,

\begin{align}
 \Phi=\frac{1}{N_c}\text{Tr}L,~~\bar{\Phi}=\frac{1}{N_c}\text{Tr}L^{\dagger}.
 \label{equ2}
\end{align}
This trace is in the color space. The Polyakov loop operator $L$ is the Wilson loop in the temporal direction 
which can be expressed as \cite{baym,megias}, 
\begin{align}
 L(\vec{x})=\mathcal{P}\exp\bigg[i\int_0^{\beta}d\tau A_0(\vec{x},\tau)\bigg], ~~\beta=1/T, 
%~~A_4=iA^0. 
 \label{equ3}
\end{align}
where, $\mathcal{P}$ is a path ordering operator in the imaginary time $\tau=it$. In a gauge where, $A^0$ is time independent,
one can perform the integration trivially and we will have $L=exp(i\beta A_0)$. Further, one can rotate the gauge field
in the Cartan sub algebra $A^c_0= A^3_0\lambda^3+ A^8_0\lambda^8$, so that $L$ is diagonal
in the color space \cite{MintzRamos}.

In the absence of quarks, the Polyakov loop can be considered as an order parameter for the confinement-deconfinement transition. Confinement-deconfinement transition 
is characterized by the spontaneous breaking of the $Z(3)$
center symmetry of QCD. For vanishing chemical potential $\Phi=\bar{\Phi}$, but at 
finite baryon chemical potential in general $\Phi\neq\bar{\Phi}$.
At low temperatures, the Polyakov loop potential has a unique minimum at $\Phi=0=\bar{\Phi}$.
However, at a higher temperature, above the transition temperature
an absolute minimum of $\mathcal{U}$ occurs at a nonvanishing value of $\Phi$ and $\bar{\Phi}$. 
For vanishing baryon chemical potential, in the high-temperature limit 
$T\rightarrow \infty$, $\Phi\rightarrow 1$. The effective potential $\mathcal{U}$ which is 
written in terms of $\Phi$  and $\bar{\Phi}$, following the $Z(3)$ symmetry is 
expressed as,  

\begin{align}
 \mathcal{U}(\Phi,\bar{\Phi},T)=\bigg[-\frac{b_2(T)}{2}\bar{\Phi}\Phi-\frac{b_3}{6}\left(\Phi^3+\bar{\Phi}^3\right)+\frac{b_4}{4}(\Phi\bar{\Phi})^2\bigg]T^4,
 \label{equ4}
\end{align}

with,

\begin{align}
 b_2(T)=a_0+a_1\left(\frac{T_0}{T}\right)+a_2\left(\frac{T_0}{T}\right)^2+a_3\left(\frac{T_0}{T}\right)^3.
 \label{equ5}
\end{align}

The coefficients $a_i$ and $b_i$ and $T_0$ can be fixed using the pure-gauge Lattice 
QCD data. The coefficients $a_i$ and $b_i$ are given in Table 1. The critical 
temperature $T_0$ for confinement-deconfinement phase transition is
fixed to be 270 MeV in the pure gauge sector \cite{rattiPNJL1,rattiPNJL2,rattiPNJL3,rattiPNJL4}. 

\begin{table}[h!]
\centering
\caption{Parameters for Polyakov loop potential }
\label{tab:table1}
\begin{tabular}{c|c|c|c|c|c}
$a_0$ & $a_1$ & $a_2$ & $a_3$ & $b_3$ & $b_4$\\
    \hline
6.75 & -1.95 & 2.625 & -7.44 & 0.75 & 7.5
\end{tabular}
\end{table}

In the mean-field approximation 
the thermodynamic potential ($\Omega_{})$ of the PNJL model at finite temperature ($T$) and quark chemical potential ($\mu$) 
can be expressed as \cite{sasakiPNJL1,NJLcpviolation2},
\begin{align}
 \Omega_{}(\sigma,\eta,\Phi,\bar{\Phi},a,T,\mu)=\Omega_q(\sigma,\eta,a,\Phi,\bar{\Phi},T,\mu)-g_2(\eta^2-\sigma^2)~\cos\left(\frac{a}{f_a}\right)+g_1(\eta^2+\sigma^2)-2g_2~\sigma\eta~\sin\left(\frac{a}{f_a}\right)+\mathcal{U}(\Phi,\bar{\Phi},T),
 \label{equ6}
\end{align}
here $\sigma = \langle\bar{q}q\rangle$ is the scalar condensate and 
$\eta=\langle\bar{q}i\gamma_5q\rangle$ is the pseudoscalar condensate. 
In the mean field approximation the fields are replaced by their expectation (thermal expectation) values. Throughout the manuscript we have used the notation $\sigma$, $\eta$, $\Phi$ and $\bar{\Phi}$ to represent the fields as well as their expectation values for simplicity and convenience. As mentioned above at a low temperature where $\Phi = 0$ and $\bar{\Phi} =0$, the potential has only one minimum. For temperatures higher than the transition temperature, the $\Phi$ and $\bar{\Phi}$ develop a nonvanishing vacuum expectation value, and the cubic term in the Polyakov loop potential leads to $Z(3)$ degenerate vacua. 
Also note that 
we are considering vacuum to be isospin symmetric. 
Generally for $a=0$ or $\theta=0$ the pseudoscalar condensate $\eta$ vanishes. Therefore parity as well as CP is not spontaneously broken at $\theta=0$. But any nonvanishing value of $\eta$ indicates the spontaneous breaking of parity symmetry and spontaneous CP violation. Any nonvanishing $\eta$ emphasizes the fact that the axion field couples to the axial current. In the PNJL model the fermionic contribution to the thermodynamic 
potential in the grand canonical ensemble is,

\begin{align}
 \Omega_q & =-4 N_c\int\frac{d^3p}{(2\pi)^3}E_p-4T\int\frac{d^3p}{(2\pi)^3}\bigg(\log\bigg[1+3\Phi\exp[-\beta(E_p-\mu)]+3\bar{\Phi}\exp[-2\beta(E_p-\mu)]+\exp[-3\beta(E_p-\mu)]\bigg]\nonumber\\
 &~~~~~~~~~~~~~~~~~~~~ +\log\bigg[1+3\bar{\Phi}\exp[-\beta(E_p+\mu)]+3\Phi\exp[-2\beta(E_p+\mu)]+\exp[-3\beta(E_p+\mu)]\bigg]\bigg),
 \label{equ7}
\end{align}
here $N_c=3$ is the number of colors, $\beta=1/T$ is the inverse of temperature and the single-particle energy $E_p$ and the effective mass are expressed as, 

\begin{align}
 E_p=\sqrt{p^2+M^2},~~M=\sqrt{(m+\alpha_0)^2+\beta_0^2}.
\label{eq8}
\end{align}

The scalar and pseudoscalar condensate $\sigma$ and $\eta$ respectively enters in to 
the expression of effective mass $M$ and single-particle energy $E_p$ through the 
functions $\alpha_0$ and $\beta_0$ which are given as\cite{NJLaxion1}, 

\begin{align}
 & \alpha_0=-2\left(g_1+g_2\cos\left(\frac{a}{f_a}\right)\right)\sigma+
2g_2~\eta~\sin\left(\frac{a}{f_a}\right)
\label{eq9}
\end{align}

\begin{align}
 & \beta_0=-2\left(g_1-g_2\cos\left(\frac{a}{f_a}\right)\right)\eta+
2g_2~\sigma~\sin\left(\frac{a}{f_a}\right).
\label{eq10}
 \end{align}

 The fermionic contribution to the thermodynamic potential of the PNJL model as given in Eq.\eqref{equ7} involves 
a vacuum contribution $(T=0,\mu=0)$ and a medium contribution $(T\neq0,\mu\neq0)$. The vacuum term is ultraviolet (UV) divergent. Various regularization methods have been used in literature to regulate this vacuum term, e.g. sharp three momentum cutoff, medium separation regularization scheme, proper time regularization scheme, etc. In this investigation we consider a sharp three momentum cutoff ($\Lambda$) to regulate the vacuum term. In the medium-term distribution function takes care of the ultraviolet problem. The parameters used in the NJL part are the same as in Refs.
\cite{NJLaxion1,frank,NJLcpviolation2}.
Here the cutoff parameter $\Lambda=590$ MeV and the bare quark mass is taken as $m=6$ MeV.
Here $g_1=(1-c)g$ and $g_2=cg$, where $c=0.2$ and $g=2.435/{\Lambda}^2$ \cite{NJLcpviolation2,NJLcpviolation5}. Parameter $c$ determines the effect of instantons and $0 \leq c \leq 0.5$ \cite{NJLcpviolation2,NJLcpviolation5}. At $\theta = 0$ the quark scalar condensate is only determined by $\Lambda, m $ and the combination $g=g_1
+g_2$. These parameters 
are fixed by fitting the physical pion mass $m_{\pi}=140.2$ MeV and pion decay constant 
$f_{\pi}=92.6$ MeV. Some comment about the choice of the parameter $c$ is in order here.  The parameter $c$
can be fixed from the mass of the isoscalar
pseudoscalar particle that arises  in the spectrum from the breaking of $U(1)_A$ axial symmetry. In a two flavor case, this
isoscalar
pseudoscalar meson can be identified with the $\eta$ meson. The mass of $\eta$ 
meson can be approximately written in terms of pion mass $(\pi)$ and the constituent quark mass $(M)$ \cite{Dmitrasinovic1996}, 
\begin{align}
m_{\eta}^{2}=m_{\pi}^{2}+\frac{g_{2} M^{2}}{\left(g_{1}^{2}-g_{2}^{2}\right) f_{\pi}^{2}}.
\label{neweq11}
\end{align}
 From Eq.\eqref{neweq11} it is clear that for $g_1=g_2$ i.e. for $c=0.5$ the $\eta$ meson disappears from the spectrum. However with the physical mass of the $\eta$ meson, i.e. $m_{\eta}= 547.8$ MeV, Eq.\eqref{neweq11} leads to a value of $c \simeq 0.09$ \cite{hmdeepak}. On the other hand, for a realistic description of $\eta$ meson, i.e. for a better way to fix the parameter $c$, one should consider a three flavour NJL model including the strange quarks. For three flavor NJL model the determinant interaction becomes a six fermion
interaction which leads to $\eta-\eta^{\prime}$ splitting. From the analysis of the $\eta-\eta^{\prime}$
splitting in the three-flavor NJL model predicts that $c\sim 0.2$ is favorable \cite{BuballaReview,frank}. It should be pointed out that even in such cases, the value of $c$ can vary
about 25\% to 30 \% (i.e., from $c\sim 0.21$ to $c\sim 0.16$) depending upon the different parametrization considered in NJL model. In the present investigation we consider $c = 0.2$ in accordance with Refs. \cite{NJLcpviolation2,NJLcpviolation5,NJLaxion1,NJLaxion2,sasakiPNJL1}. 
 
 The physical values of the condensates $\sigma_0,\eta_0,\Phi_0$ and $\bar{\Phi}_0$ can 
be obtained by solving the gap equations,

 \begin{align}
  \frac{\partial\Omega_{}}{\partial\sigma}=0;~~  \frac{\partial\Omega_{}}{\partial\eta}=0;~~ \frac{\partial\Omega_{}}{\partial\Phi}=0;~~
   \frac{\partial\Omega_{}}{\partial\bar{\Phi}}=0.
   \label{equ11}
 \end{align}

Note that the physical values of the condensates, $\sigma_0,\eta_0,\Phi_0$ and $\bar{\Phi}_0$ are functions of $a,T$ and $\mu$.
The effective thermodynamic
potential for the QCD axion within the framework of PNJL model in a hot and dense medium  is then given by,

\begin{align}
\tilde{\Omega}(a,T,\mu)=\Omega\bigg[\sigma_0(a,T,\mu),\eta_0(a,T,\mu),\Phi_0(a,T,\mu),\bar{\Phi}_0(a,T,\mu),a,T,\mu\bigg]. 
\label{equ12}
\end{align}

Using the axion potential one can obtain the axion mass and the axion self coupling  can be obtained as,

\begin{align}
 m_a^2=\frac{d^2\tilde{\Omega}}{da^2}|_{a=0}=\frac{\chi}{f_a^2};~~~\lambda_a=\frac{d^4\tilde{\Omega}}{da^4}|_{a=0}.
 \label{equ13}
\end{align}

Here $\chi$ is the topological susceptibility. Note that all the physical condensates $\sigma_0$, $\eta_0$, $\Phi_0$ and $\bar{\Phi}_0$ have implicit dependence on the axion field. Hence,

\begin{align}
 \frac{d\tilde{\Omega}}{da}=\frac{\partial\tilde{\Omega}}{\partial a}+\frac{\partial\tilde{\Omega}}{\partial \sigma}\frac{\partial\sigma}{\partial a}+\frac{\partial\tilde{\Omega}}{\partial \eta}\frac{\partial\eta}{\partial a}+\frac{\partial\tilde{\Omega}}{\partial \Phi}\frac{\partial\Phi}{\partial a}+\frac{\partial\tilde{\Omega}}{\partial \bar{\Phi}}\frac{\partial\bar{\Phi}}{\partial a}.
 \label{equ14}
\end{align}

Therefore to evaluate axion mass and self coupling as given in Eq.\eqref{equ13} we have to evaluate $\frac{\partial^{(n)}\sigma}{\partial a^{(n)}}$, $\frac{\partial^{(n)}\eta}{\partial a^{(n)}}$, $\frac{\partial^{(n)}\Phi}{\partial a^{(n)}}$ and $\frac{\partial^{(n)}\bar{\Phi}}{\partial a^{(n)}}$, where $n=1,2,3,4$ represents the order of the derivative. $\frac{\partial^{(n)}\sigma}{\partial a^{(n)}}$, $\frac{\partial^{(n)}\eta}{\partial a^{(n)}}$, $\frac{\partial^{(n)}\Phi}{\partial a^{(n)}}$ and $\frac{\partial^{(n)}\bar{\Phi}}{\partial a^{(n)}}$ can be evaluated by taking successive derivative of the gap equations 
as given in Eq.\eqref{equ11}, with respect to the axion field $a$. 

\section{RESULTS AND DISCUSSION}
\label{results}

\subsection{Vanishing quark chemical potential}

\begin{figure}
%\centering
\begin{minipage}{.5\textwidth}
  \centering
  \includegraphics[width=1.15\linewidth]{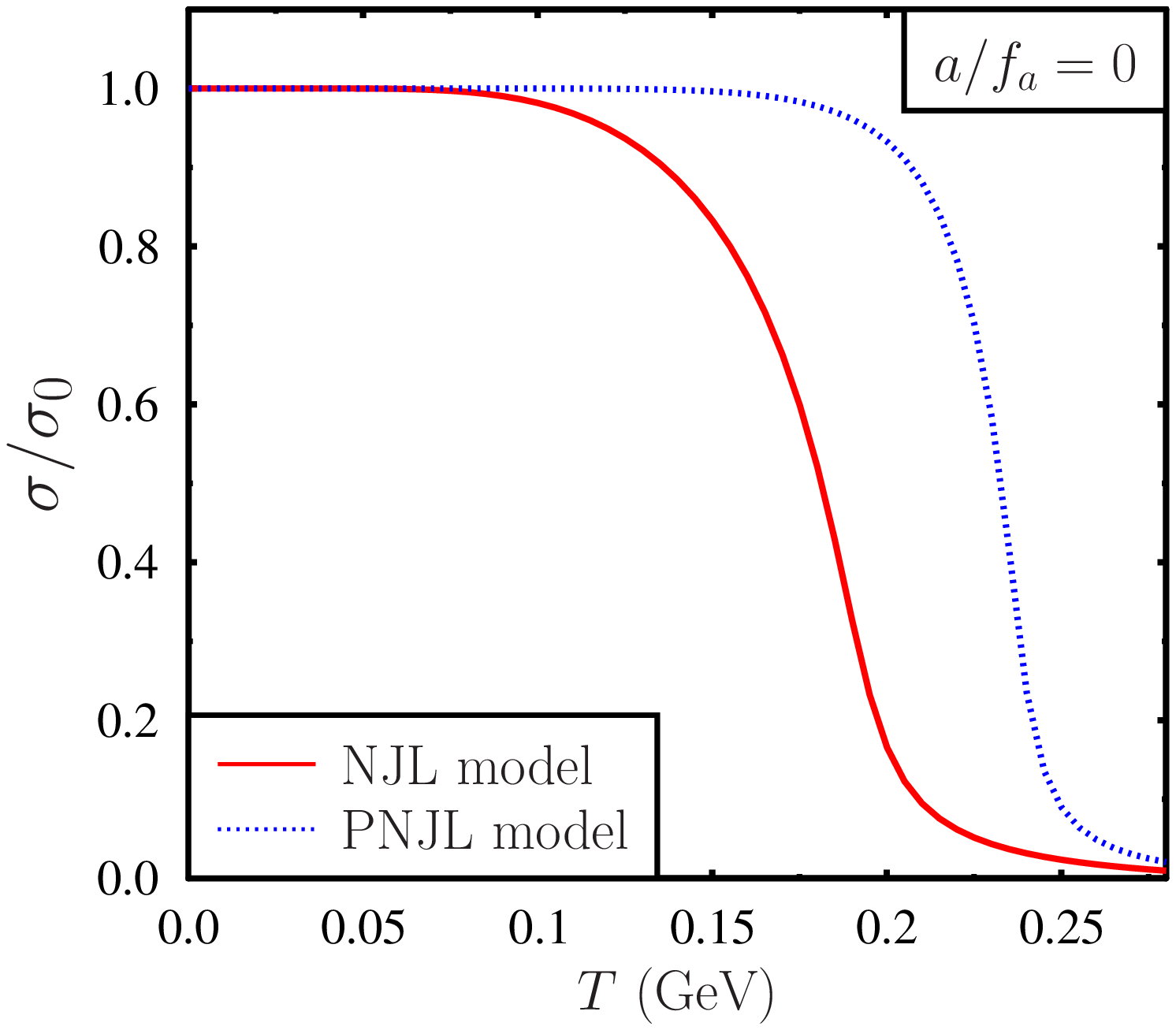}
\end{minipage}%
\begin{minipage}{.5\textwidth}
  \centering
  \includegraphics[width=1.15\linewidth]{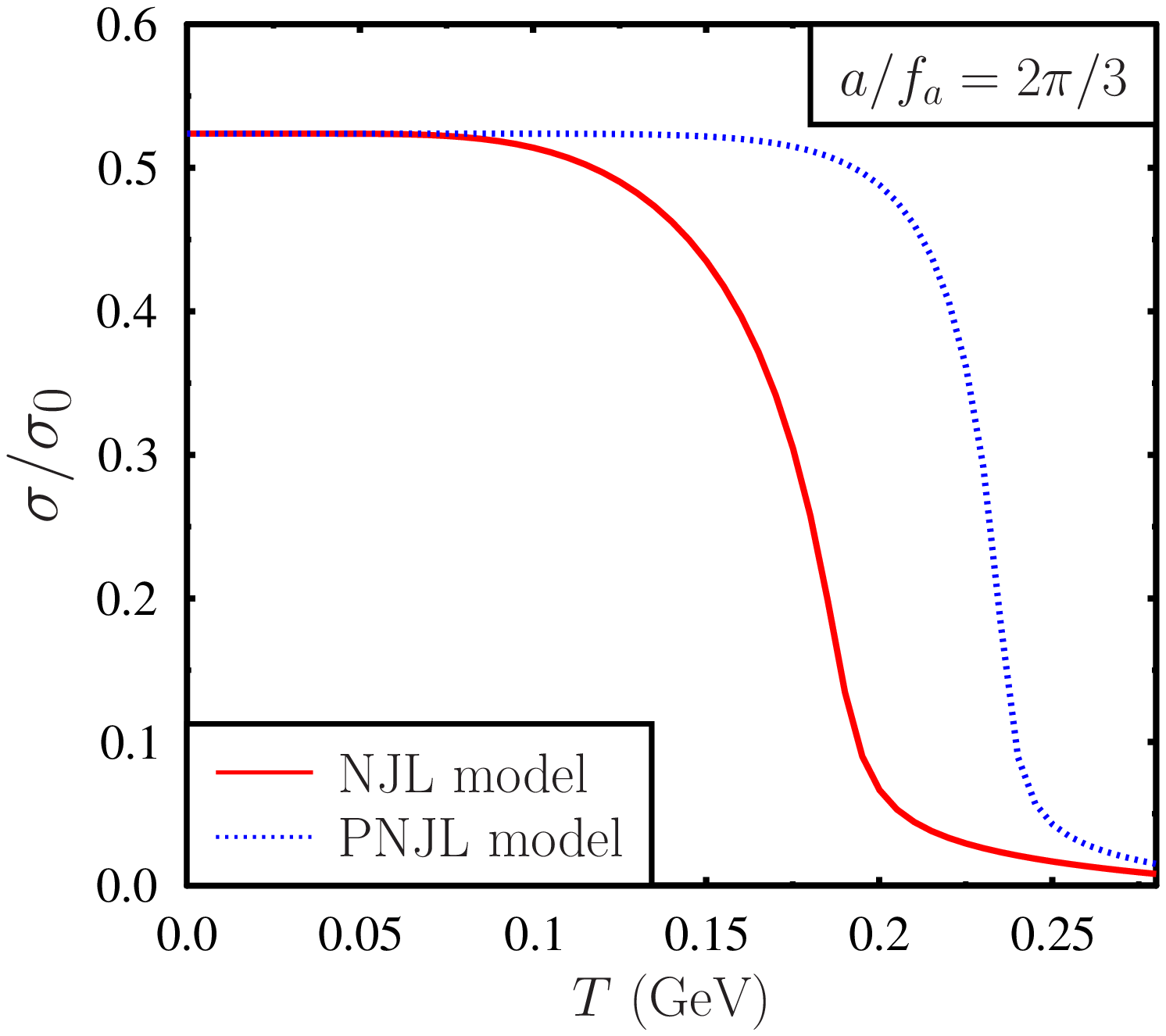}
\end{minipage}
\caption{Left plot: variation of the normalized chiral condensate $\sigma/\sigma_0$ with
temperature for $a/f_a=0$. Right plot: variation of the normalized chiral condensate $\sigma/\sigma_0$ with
temperature for $a/f_a=2\pi/3$. Here $\sigma_0$ denotes the scalar condensate at in vacuum for $a/f_a=0$.}
\label{Fig.1}
\end{figure}

\begin{figure}
\centering
\begin{minipage}{.5\textwidth}
  \centering
  \includegraphics[width=1.15\linewidth]{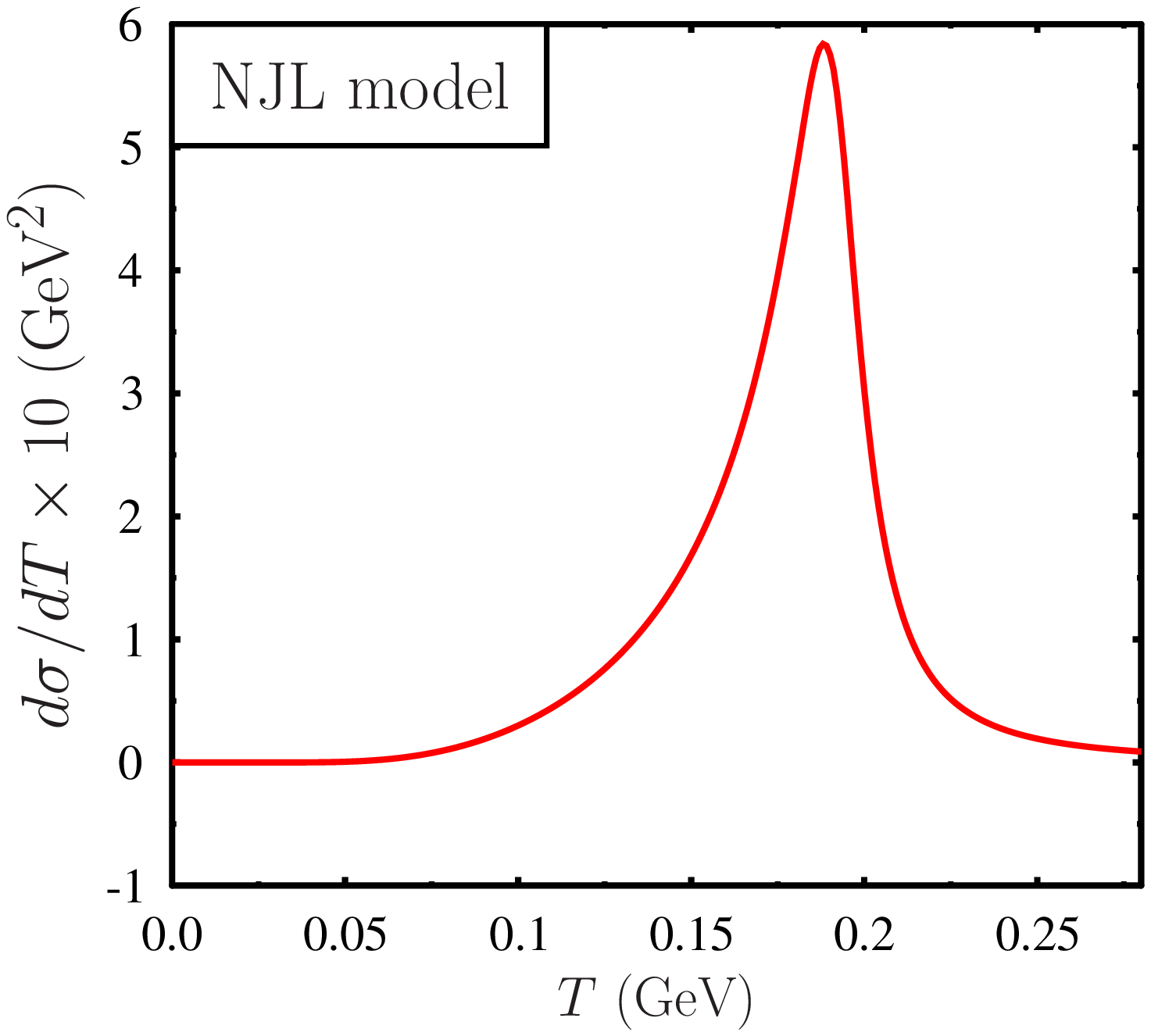}
\end{minipage}%
\begin{minipage}{.5\textwidth}
  \centering
  \includegraphics[width=1.15\linewidth]{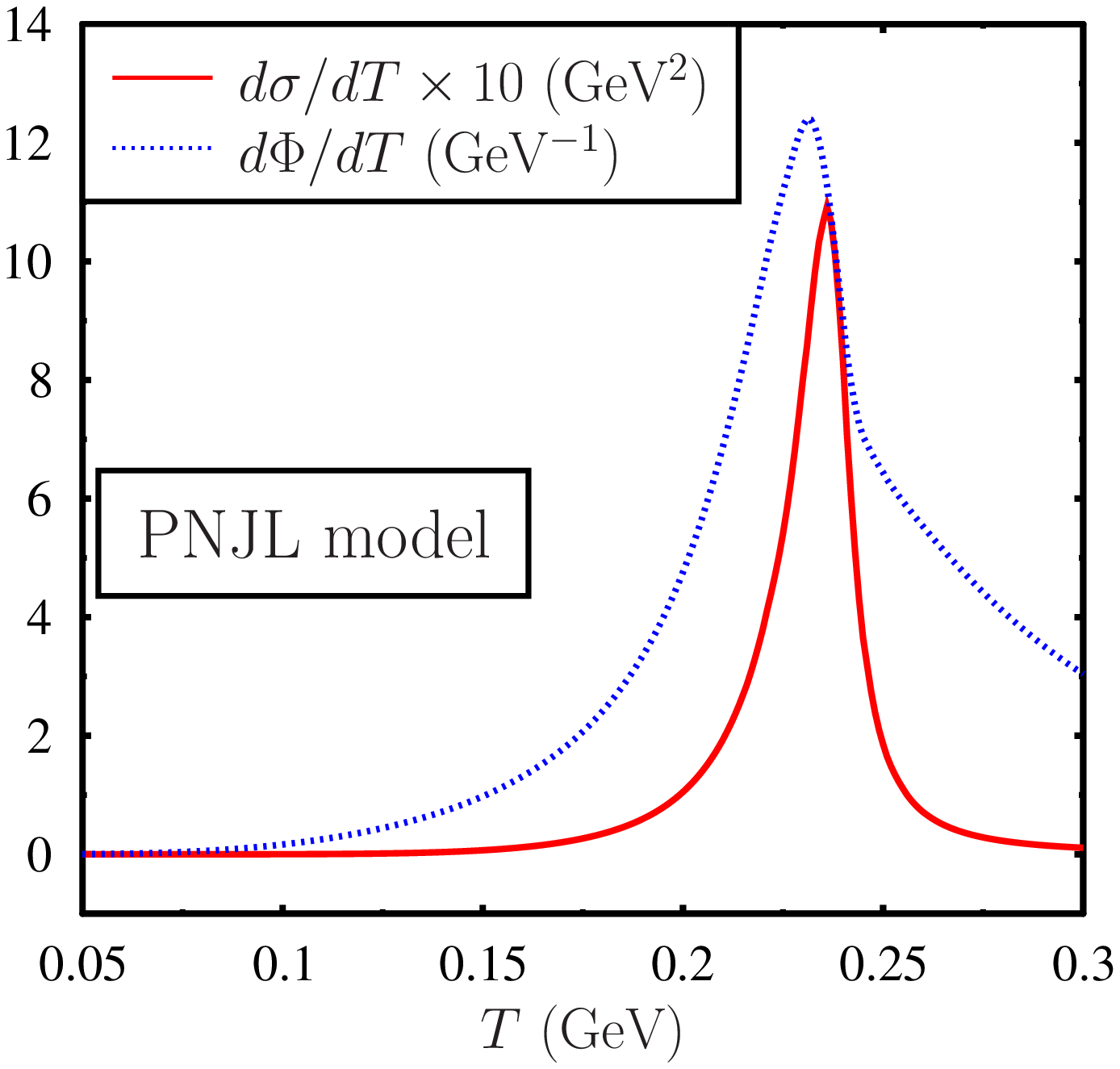}
\end{minipage}
\caption{Left plot: variation of $d\sigma/dT$ with temperature in the NJL model. Peak structure in the variation of $d\sigma/dT$ indicates that the location of the pseudo critical temperature in the NJL model which is $T_c\sim 190$ MeV for the parameter set considered here. Right plot: variation of $d\sigma/dT$ and $d\Phi/dT$ with temperature in the PNJL model. From this figure we can see that the pseudo critical temperatures for the chiral transition and the confinement-deconfinement transition are almost same , i.e. chiral transition and the confinement-deconfinement occurs almost simultaneously.}
\label{suscwt}
\end{figure}

\begin{figure}
    %\centering
    \includegraphics[width=0.7\textwidth]{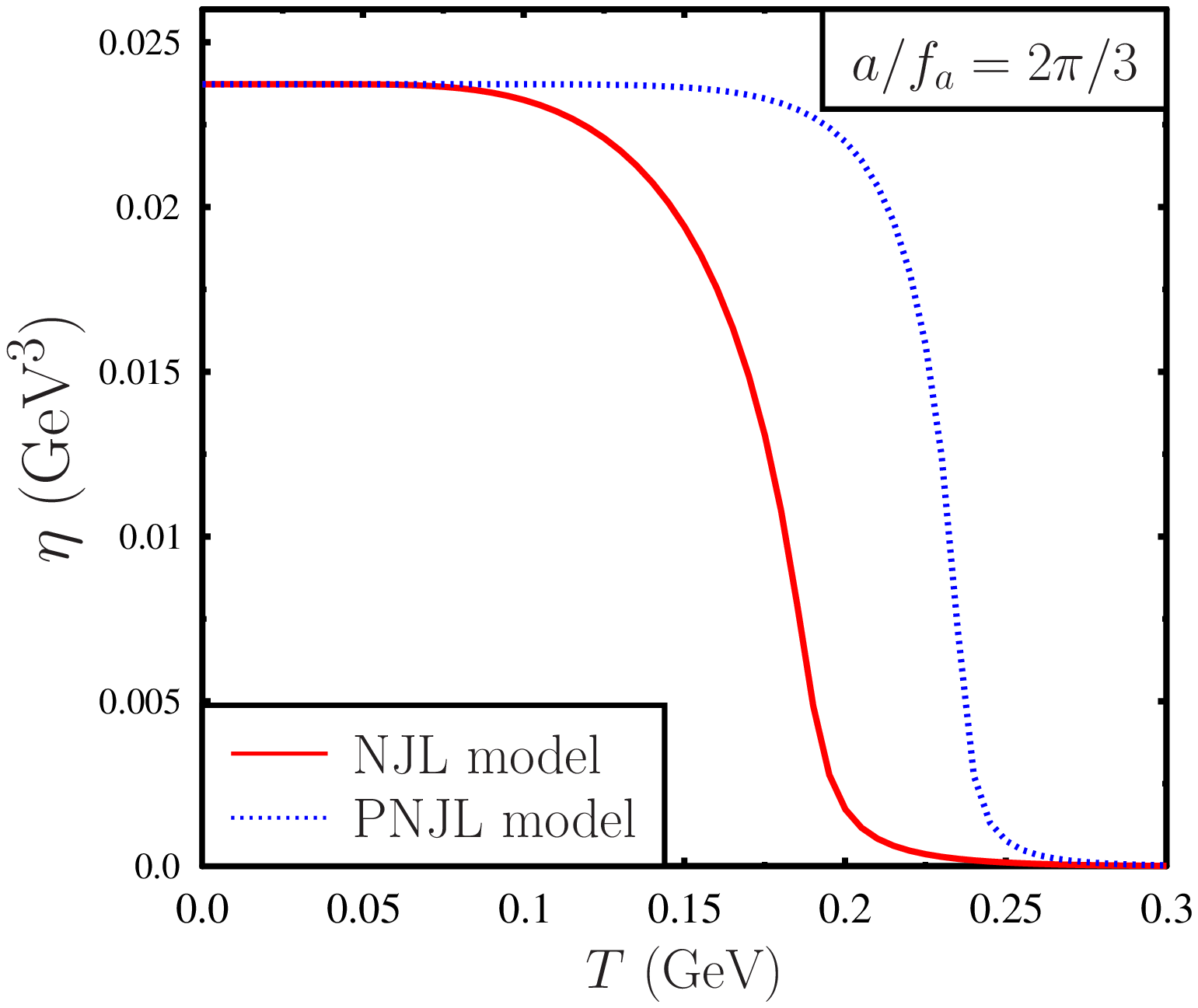}
    \caption{Variation of the pseudo scalar condensate ($\eta$) with temperature for $a/f_a=2\pi/3$ in the NJL and PNJL model. Non vanishing value of the pseudo scalar condensate indicates the spontaneous breaking of the parity symmetry (P) or equivalently the breaking of the CP (charge conjugation and parity) symmetry in QCD. At zero temperature CP symmetry is spontaneously broken for $a/f_a=2\pi/3$, but at a high temperature this symmetry is restored. In PNJL model restoration of the CP symmetry happens at a relatively higher temperature as compared to the NJL model.}
    \label{Fig.2}
\end{figure}

\begin{figure}
    %\centering
    \includegraphics[width=0.7\textwidth]{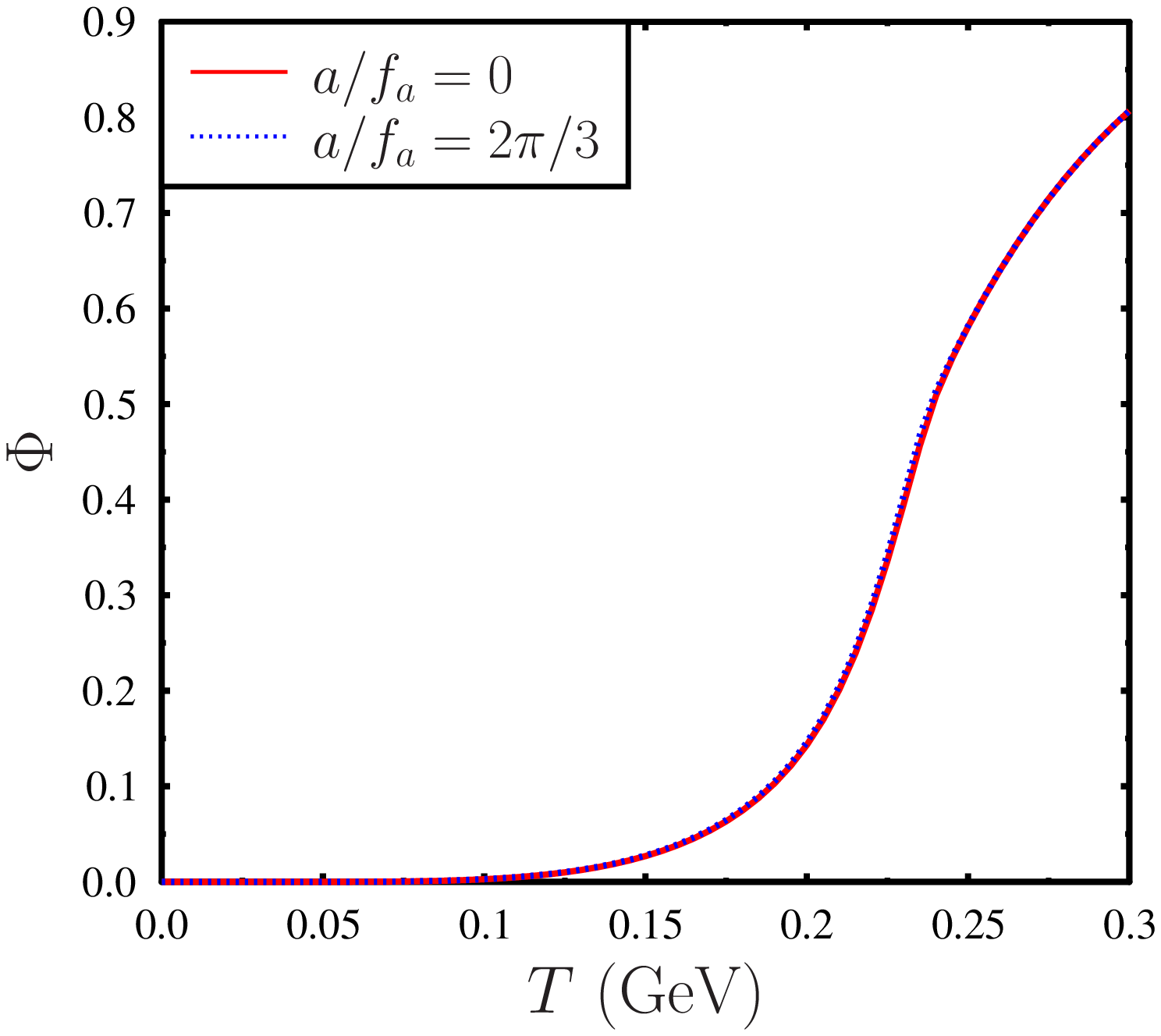}
    \caption{Variation of the Polyakov loop $\Phi$ with temperature for $a/f_a=0$ and $a/f_a=2\pi/3$. For vanishing quark chemical potential $\Phi$ and $\bar{\Phi}$ are same. From this plot we can observe that for the low temperature range $T\lesssim 0.1$ GeV, $\Phi$ is not large. Only for a high temperature range $\Phi$ is significant. Due to the absence of any direct interaction between the Polyakov loop and the axion field in the PNJL model, variation of $\Phi$ is not strongly affected by the axion field.}
    \label{Fig.3}
\end{figure}

In this subsection, we present the results for vanishing quark chemical potential $\mu=0$. The Polyakov loop $\Phi$
and its conjugate $\bar{\Phi}$ are the same at zero chemical potential. 
 In general, the fields $\Phi$ and $\bar{\Phi}$ are different at non-zero quark chemical potential. From the thermodynamic potential of the PNJL model as given in Eq.\eqref{equ6}, we can observe that the condensates $\Phi$ and $\bar{\Phi}$ only appears at finite temperature. Therefore the presence of $\Phi$ and $\bar{\Phi}$ can only affect the other condensates at finite temperature. The effect of the Polyakov loop becomes significant around the transition temperature. In fact due to the nonvanishing values of $\Phi$ and $\bar{\Phi}$ the chiral transition temperature in the PNJL model changes to a higher value with respect to the same in the NJL model.
 The NJL model results are obtained from Eq.\eqref{equ6} and Eq.\eqref{equ7} by replacing $\Phi = \bar{\Phi}=1$ in Eq.\eqref{equ7} and taking $\mathcal{U}(\Phi,\bar{\Phi},T)=0$ in Eq.\eqref{equ6}. Thus we are taking the same parameters in the NJL and PNJL model in the quark sector.
 All these features of having a nonvanishing value of $\Phi$ and $\bar{\Phi}$ have been demonstrated in Fig. \eqref{Fig.1}, Fig.\eqref{suscwt},  Fig.\eqref{Fig.2} and Fig.\eqref{Fig.3} .
 In Fig.\ref{Fig.1} we show the variation of the normalized chiral 
condensate $\sigma/\sigma_0$ with temperature in NJL and PNJL model for two different values of $a/f_a$. Here $\sigma_0$ is the chiral condensate in vacuum for $a/f_a = 0$.  Left plot in Fig.\ref{Fig.1} is for
$a/f_a=0$ and the right plot in Fig.\ref{Fig.1} is for $a/f_a=2\pi/3$. From Fig.\ref{Fig.1}, it is clear that 
the normalized condensate is same in both NJL and PNJL model at 
low temperature ($\sim$ 100 MeV). This is because the Polyakov loop is very 
small ($\sim 0$) at low temperature upto 100 MeV which is clearly seen 
in Fig.\ref{Fig.3}. However, the Polyakov loop becomes nonzero after this 
temperature, hence the normalized condensate differs in NJL and PNJL model at a higher temperature.

The normalized 
condensate $\sigma/\sigma_0$ in the NJL model starts to decrease from its maximum value at a relatively lower temperature as compared to the PNJL model. The normalized condensate starts to drop at a higher temperature ($ T \sim 200$ MeV)
in the PNJL model as compared to the NJL model, where $\sigma/\sigma_0$ starts to decrease at a relatively lower temperature ($ T\sim 100$ MeV).
This is because in PNJL model the chiral transition temperature is higher with respect to the same in the NJL model, which has been clearly shown in Fig.\eqref{suscwt}. In the left plot in Fig.\eqref{suscwt} we show the variation of $d\sigma/dT$ with temperature for the NJL model. In the right plot in Fig.\eqref{suscwt} we show the variation of $d\sigma/dT$ and $d\Phi/dT$ with temperature for the PNJL model. From Fig.\eqref{suscwt} we can see that the variation of $d\sigma/dT$ and $d\Phi/dT$ shows a nonmonotonic behavior with temperature with a peak. The peak in the variation of $d\sigma/dT$ and $d\Phi/dT$ indicates the chiral transition temperature and the confinement-deconfinement transition temperature respectively. It is clear from Fig.\eqref{suscwt} that the chiral transition temperature is higher in the PNJL model as compared to the NJL model. In other words, one can say that chiral symmetry is restored at a lower temperature in the NJL model compared to the PNJL model. This is also true for $a/f_a=2\pi/3$
as shown in the right plot of Fig.\ref{Fig.1}. The normalized condensate value is smaller for $a/f_a=2\pi/3$
compared to $a/f_a=0$ case. Also note that in the PNJL model the two transition temperatures, i.e. the chiral transition temperature and the confinement-deconfinement transition temperature are exactly not the same but they are very close. The left plot in Fig.\ref{suscwt}
shows the chiral transition temperature $T_c \sim 190$ MeV in NJL
model. But, the chiral transition temperature is 235 MeV and the deconfinement transition temperature is 230 MeV in the PNJL model. So, we have taken the pseudo critical temperature,  $T_c = 230$ MeV in the PNJL model.

We have shown the variation of $\eta$ with temperature in Fig.\ref{Fig.2}
for $a/f_a=2\pi/3$. Note that $\eta$ is always zero for $a/f_a=0$. $\eta$ has a similar kind 
of variation as the condensate $\sigma$ as shown in Fig.\ref{Fig.1}. For a nonvanishing value of $a/f_a$, $\eta$ condensate ($\eta\neq 0$) occurs at low temperature and hence the parity symmetry is spontaneously broken. At higher temperature $T\sim 200$ MeV $\eta$ is close to zero in NJL model. Therefore the parity symmetry is restored at a higher temperature. The transition temperature for the parity symmetry restoration for the nonvanishing value of $a/f_a$ is higher for the PNJL model with respect to the NJL model, analogous to chiral transition.

Fig.\ref{Fig.3} represents the variation of the Polyakov loop ($\Phi=\bar{\Phi}$) with temperature
for $a/f_a=0$ and $a/f_a=2\pi/3$. $\Phi$ is almost same for both $a/f_a=0$ and $a/f_a=2\pi/3$. Note that the condensates $\Phi$ and $\bar{\Phi}$ are not directly coupled to $a/f_a$. Also the Polyakov loop potential is independent of $a/f_a$. In the thermodynamic potential of PNJL model as given in Eq.\eqref{equ6} the dependence of $a/f_a$ only comes through the terms which are associated with $\sigma$ and $\eta$ condensates. Dependence of $\Phi$ and $\bar{\Phi}$ on the CP violating parameter $\theta\equiv a/f_a$ translates through $\sigma$ and $\eta$ condensates, which are connected with $\Phi$ and $\bar{\Phi}$ through the gap equations. For the temperature range for which $\Phi$ and $\bar{\Phi}$ becomes important, $\sigma$ and $\eta$ becomes less significant. Hence $a/f_a$ does not affect $\Phi$ and $\bar{\Phi}$ significantly.
$\Phi$ is almost zero at a lower temperature up to 100 MeV (confined phase), 
then it increases as temperature increases and goes towards the deconfined phase.
The Polyakov loop value approaches it's asymptotic value of unity only at very large temperature. 

\begin{figure}
    %\centering
    \includegraphics[width=0.7\textwidth]{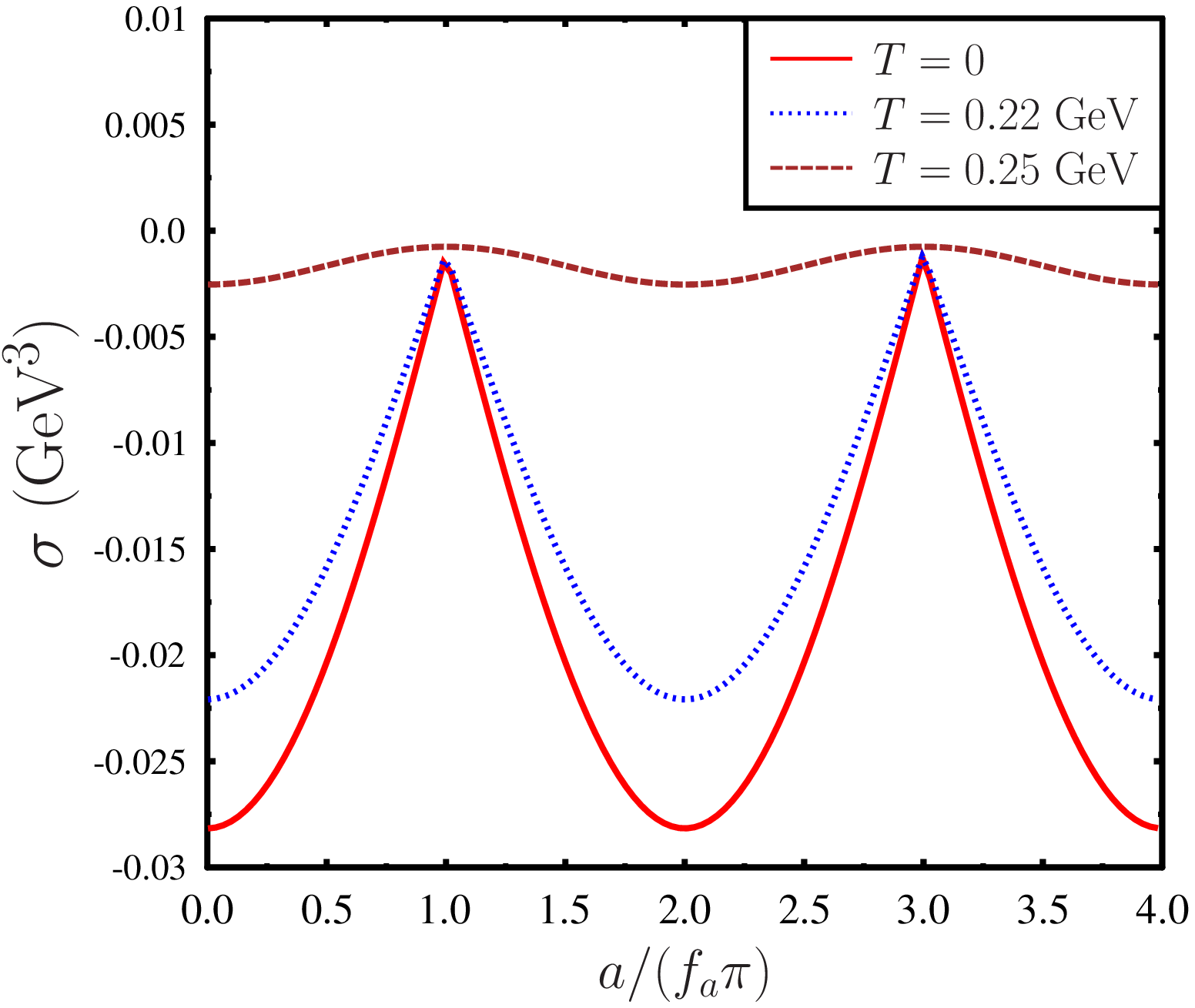}
    \caption{Variation of scalar condensate $\sigma$ with  $a/f_a$ for three different values of temperature in the PNJL model. The periodic variation of $\sigma$ with $a/f_a$ is due to the dependence of the thermodynamic potential on the periodic functions $\sin(a/f_a)$ and $\cos(a/f_a)$.}
    \label{Fig.4}
\end{figure}

Next in Fig.\eqref{Fig.4} we have shown the variation of $\sigma$ with respect to $a/f_a$ at three different 
temperatures i.e T=0, 220 and 250 MeV. The periodic  behavior of the condensate 
with respect to $a/f_a$ is due to cos($a/f_a$)
and sin($a/f_a$) terms present in the thermodynamic potential. This behavior of the condensate with $a/f_a$ 
is similar to the results as obtained in NJL model in Ref. \cite{NJLaxion1,NJLaxion2}. $\sigma$ reaches its maximum value for 
$a/f_a = (2i+1)\pi$, for i = 0, 1, 2 ...etc and attains minimum for $a/f_a = 2i\pi$ at all
temperatures.
  
\begin{figure}
    %\centering
    \includegraphics[width=0.7\textwidth]{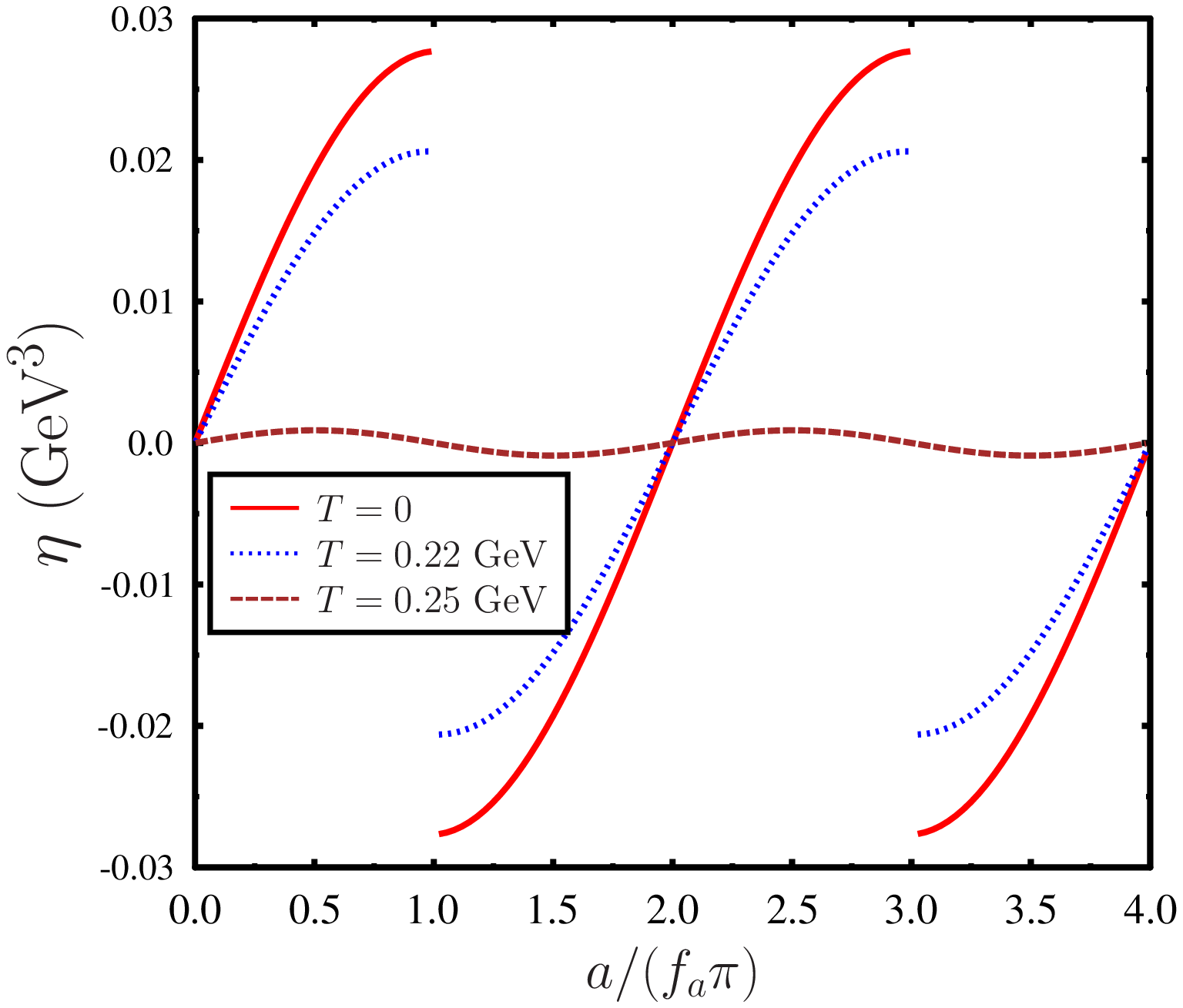}
    \caption{Variation of $\eta$ with respect to $a/f_a$ for different values of T in the PNJL model. At zero temperature $\eta$ has discontinuity at $a/f_a=(2i+1)\pi$ and $\eta$ vanishes at $a/f_a=2i\pi$, for $i=0,1,2...$. Presence of non vanishing $\eta$ indicates spontaneous breaking of the CP symmetry. The degenerate vacua present at $a/f_a = (2i+1)\pi$ are CP conjugate. With increasing temperature this degenerate vacuum structure goes away.}
    \label{Fig.5}
\end{figure}

Fig.\ref{Fig.5} represents the variation of $\eta$ with respect to $a/f_a$ for three 
different values of T = 0, 220 and 250 MeV.
At T = 0, $\eta$ is discontinuous for
$a/f_a = (2i+1)\pi$, for i= 0, 1, 2 ...etc. But $\eta$ vanishes for $a/f_a = 2i\pi$. 
At T = 0 and $a/f_a = (2i+1)\pi$ due to
spontaneous CP symmetry violation there exists two degenerate vacua which is in
agreement with the Dashen’s phenomena \cite{dashen}. The two vacua which have opposite signs of the condensate $\eta$ differ by a CP transformation between them. With increasing temperature, this degenerate vacuum structure vanishes. 
Fig.\ref{Fig.5} shows that Dashen's phenomena hold good at a higher temperature 
$T \sim 220$ MeV. But, at $T \sim 250 MeV$, this phenomena breaks down and  $\eta$ is a continuous
as a function $a/f_a$. This phenomenon breaks down at a much lower temperature T = 180 MeV in the NJL model compared to the PNJL model as seen in Ref.\cite{NJLaxion2}.

\begin{figure}
    %\centering
    \includegraphics[width=0.7\textwidth]{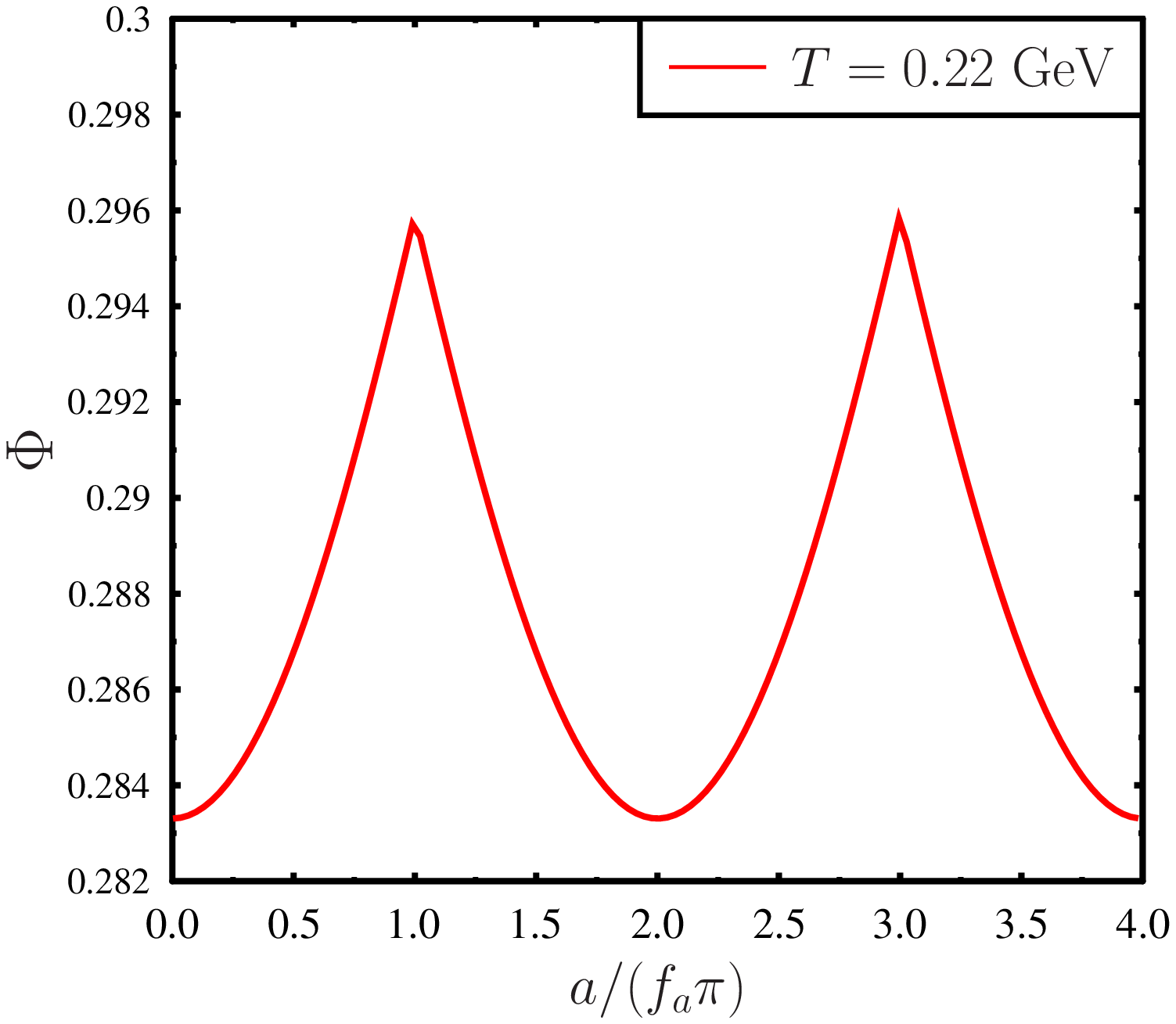}
    \caption{Variation of $\Phi$ with $a/f_a$ for T=220 MeV. Variation of $\Phi$ with $a/f_a$ is similar to the variation of scalar condensate with $a/f_a$. Note that the Polyakov loop does not have direct coupling with the axion field. However due to gap equation $\Phi$ is connected with the other condensates which depend on $a/f_a$. Therefore $\Phi$   indirectly depends on $a/f_a$.}
    \label{Fig.6}
\end{figure}

In Fig.\ref{Fig.6} we have shown the variation of Polyakov loop $\Phi$ with respect 
to $a/f_a$ at T=220 MeV. At low temperatures, $\Phi$ is almost zero. $\Phi$ becomes significant only in high-temperature range. $\Phi$ has  
similar kind of variation as the condensate $\sigma$ (as shown in Fig.\ref{Fig.4})
with respect to $a/f_a$. 
 Note that the magnitude of $\Phi$ does not change significantly with $a/f_a$. As mentioned earlier $\Phi$ does not depend upon $a/f_a$ directly, also the Polyakov loop potential is independent of $a/f_a$. However, due to the gap equations, $\Phi$ depends upon the other condensates $\sigma$ and $\eta$. Due to the variation of $\sigma$ and $\eta$ with $a/f_a$, indirectly $\Phi$ also depends on $a/f_a$. Further in the temperature range where $\Phi$ has a significantly large value, other condensates become very small. Hence the magnitude of $\Phi$ does not change much with $a/f_a$. But the dependence of $\sigma$ and $\eta$ on $a/f_a$ translates to the variation of $\Phi$ with $a/f_a$ through the gap equations.

\begin{figure}
    %\centering
    \includegraphics[width=0.7\textwidth]{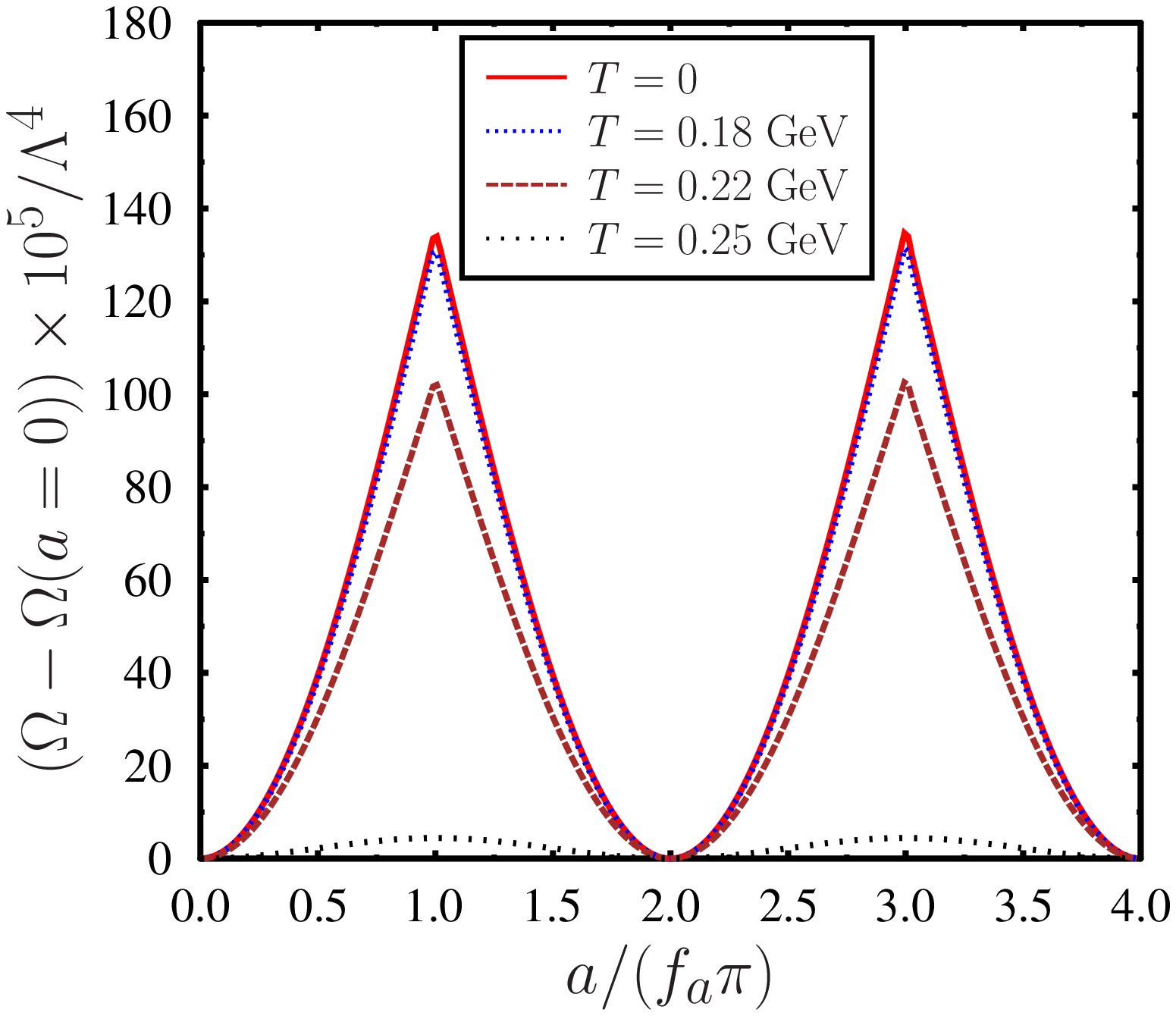}
    \caption{Variation of the normalized thermodynamic potential for different values of the temperature in the PNJL model.
The thermodynamic potential is given with respect to the
potential at $a=0$ at each temperature.}
    \label{Fig.7}
\end{figure}

Fig.\ref{Fig.7} shows the variation of the normalized thermodynamic potential or the axion potential with respect to $a/f_a$. At each temperature the value of the thermodynamic potential at $a=0$ has been subtracted.
 This potential has degenerate vacua at $a/f_a=2i\pi$ for i=0,1,2...etc. It also has maxima  
at $a/f_a=(2i+1)\pi$. According to Vafa-Witten theorem, the effective potential should have minimum at $a=0$, which is clear from this figure. At finite temperature the effective potential becomes flatter. 
As the temperature is increased from 0 to 180 MeV,
the potential does not change much. 
This is because in this temperature range the condensate values do not change much in the PNJL model as shown in Figs.\ref{Fig.1}, \ref{Fig.2} and \ref{Fig.3} respectively. Further for a higher temperature range above $180$ MeV the values of the different condensates, e.g. $\sigma,\eta$, and $\Phi$ changes significantly in PNJL model, so does the effective potential. As compared to the PNJL model in the NJL model the effective potential becomes flatter at a much lower temperature as has been reported in Ref.\cite{NJLaxion1,NJLaxion2}.  As the temperature is increased further, the barrier between degenerate vacua decreases. The barrier vanishes above $T\sim 250$ MeV.

\begin{figure}
    %\centering
    \includegraphics[width=0.6\textwidth]{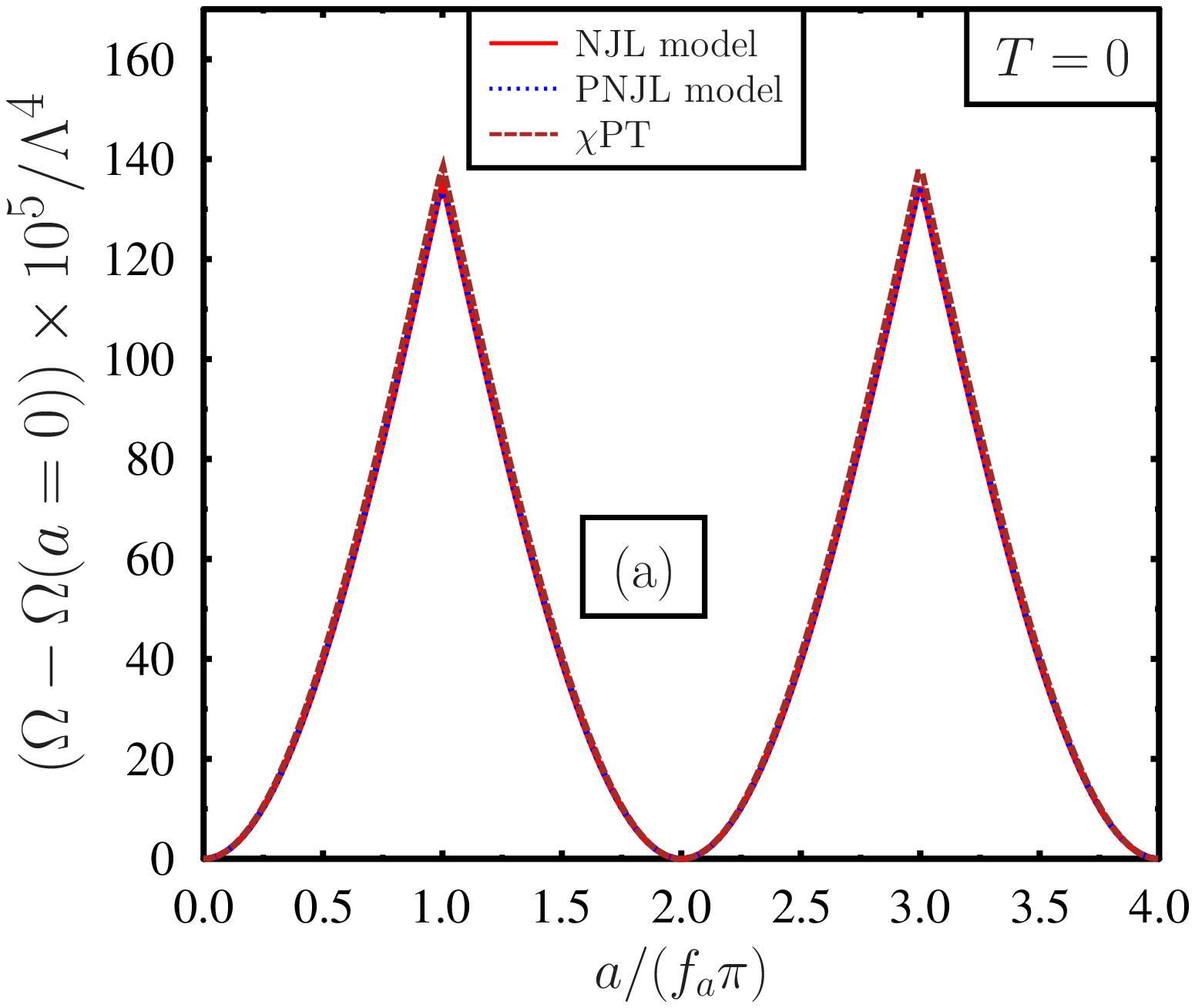}
\begin{minipage}{.5\textwidth}
  \centering
  \includegraphics[width=1.15\linewidth]{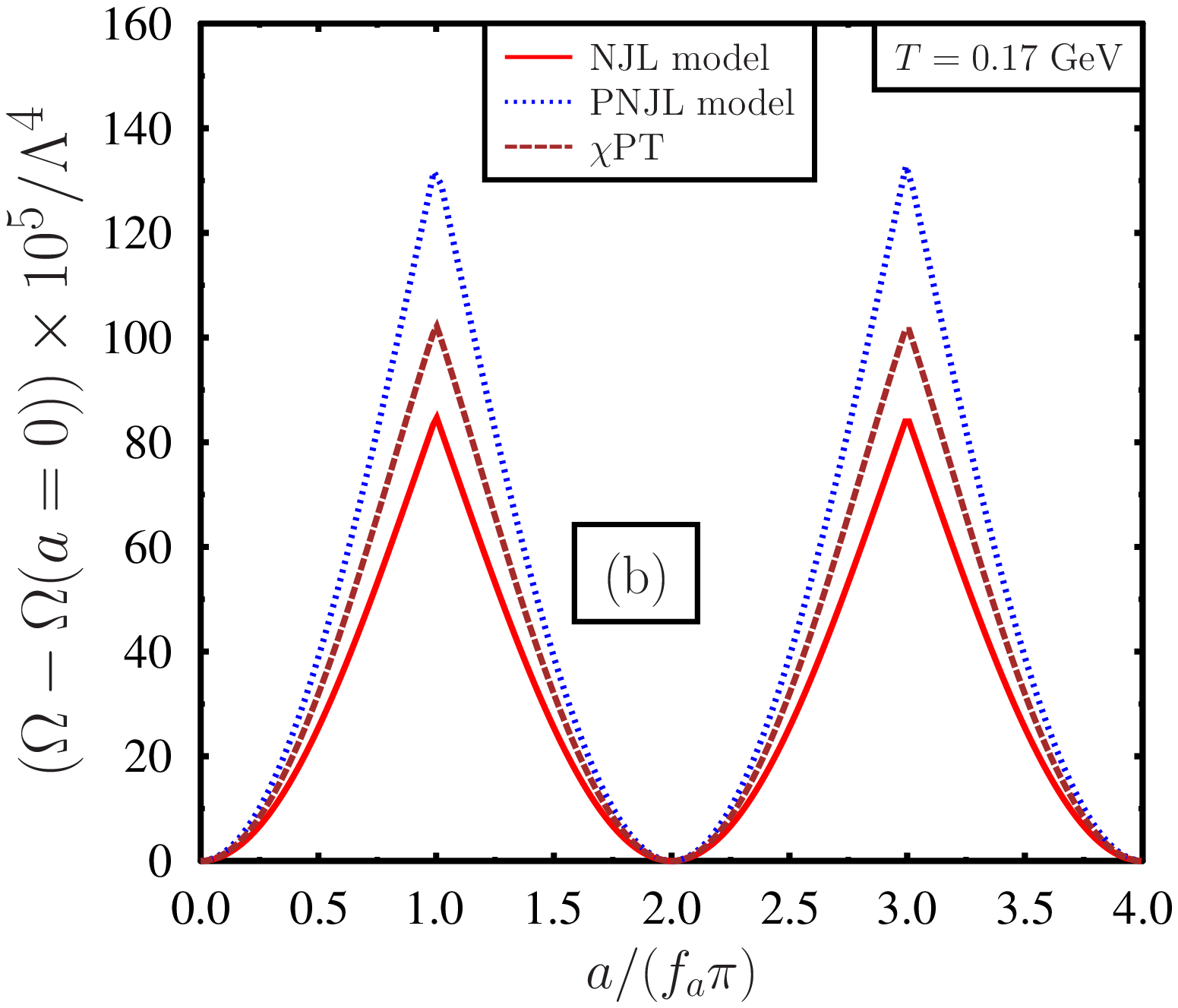}
\end{minipage}%
\begin{minipage}{.5\textwidth}
  \centering
  \includegraphics[width=1.15\linewidth]{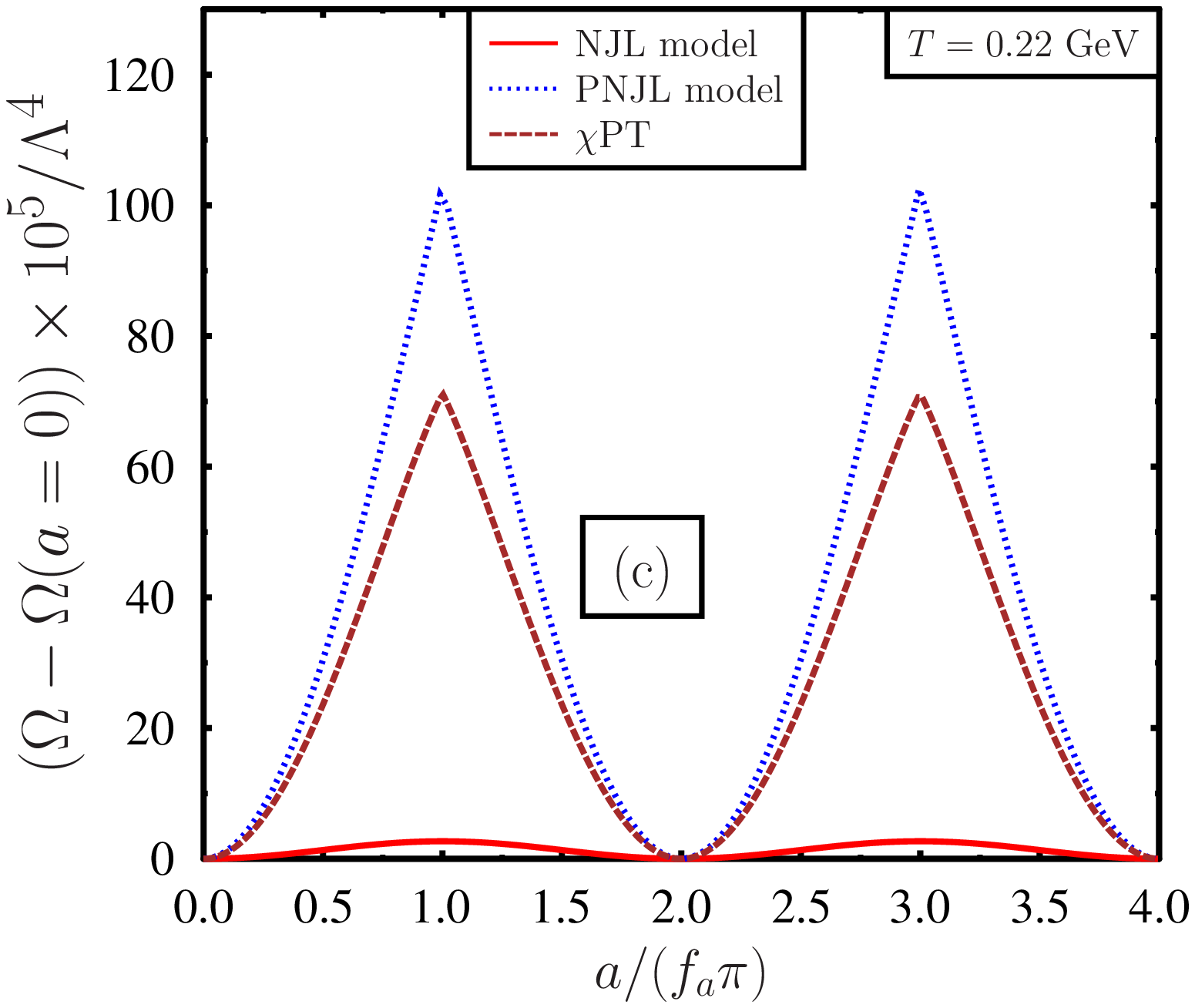}
\end{minipage}
\caption{Variation of the thermodynamic potential with $a/f_a$ in different models. Plot (a) is for zero temperature and plot (b) and plot (c) are for finite temperature. We can observe from these plots that at zero temperature all the three models (NJL, PNJL and $\chi PT$) give similar estimation of the axion contribution to the thermodynamic potential and they are equivalent. However with increasing temperature they start to differ as can be seen from plot (b) and (c).}
\label{newfig8}
\end{figure}

In Fig.\eqref{newfig8} we compare the normalized axion potential in NJL, PNJL and chiral perturbation theory $(\chi$PT). For the $SU(2)$ chiral perturbation theory up to next-to-leading order the temperature dependence of the axion potential for $m_u=m_d$ is given as \cite{NJLaxion1,chipt},

\begin{align}
 \Omega(a,T)= \Omega(a)\bigg[1-\frac{3}{2}\frac{T^4}{\pi^2f_{\pi}^2M_{\pi}^2}\times\int_0^{\infty} q^2 \log\left(1-\exp\left[-\sqrt{q^2+M_{\pi}^2/T^2}\right]\right)dq\bigg], 
\end{align}
here the zero temperature axion potential $\Omega(a)$ in the $\chi$PT model is \cite{chipt}, 
\begin{align}
 \Omega(a) = -m_{\pi}^2f_{\pi}^2\sqrt{1-\frac{4 m_u m_d}{(m_u+m_d)^2}\sin^2\left(\frac{a}{2f_a}\right)},
\end{align}
and the $a/f_a$ dependent pion mass $M_{\pi}$ can be expressed as  \cite{chipt},
\begin{align}
M_{\pi}^2\left(\frac{a}{f_a}\right) = m_{\pi}^2\sqrt{1-\frac{4 m_u m_d}{(m_u+m_d)^2}\sin^2\left(\frac{a}{2f_a}\right)}.
\end{align}
In Fig.\eqref{newfig8} the thermodynamic potential is given with respect to the
potential at $a=0$ at each temperature.
From the Fig.\eqref{newfig8}(a) we can observe that at zero temperature all the three models i.e. NJL, PNJL and the chiral perturbation theory are equivalent. On the other hand at finite temperature these models tend to differ as can be seen in Fig.\eqref{newfig8}(b) and Fig.\eqref{newfig8}(c). Note that chiral perturbation theory does not incorporate quark degree of freedom. Therefore at a high temperature in comparison with the QCD transition scale $\chi$PT predictions are not reliable. On the other hand NJL model only deals with the chiral symmetry of QCD and does not include gauge fields. Further PNJL model incorporates quark degree of freedom 
as well as temporal QCD gauge field. The Polyakov loop contribution to the thermodynamic potential becomes very important near the QCD transition scale. Therefore at finite temperature particularly near the QCD transition scale the difference in these models are expected.

\begin{figure}
\centering
\begin{minipage}{.5\textwidth}
  \centering
  \includegraphics[width=.95\linewidth]{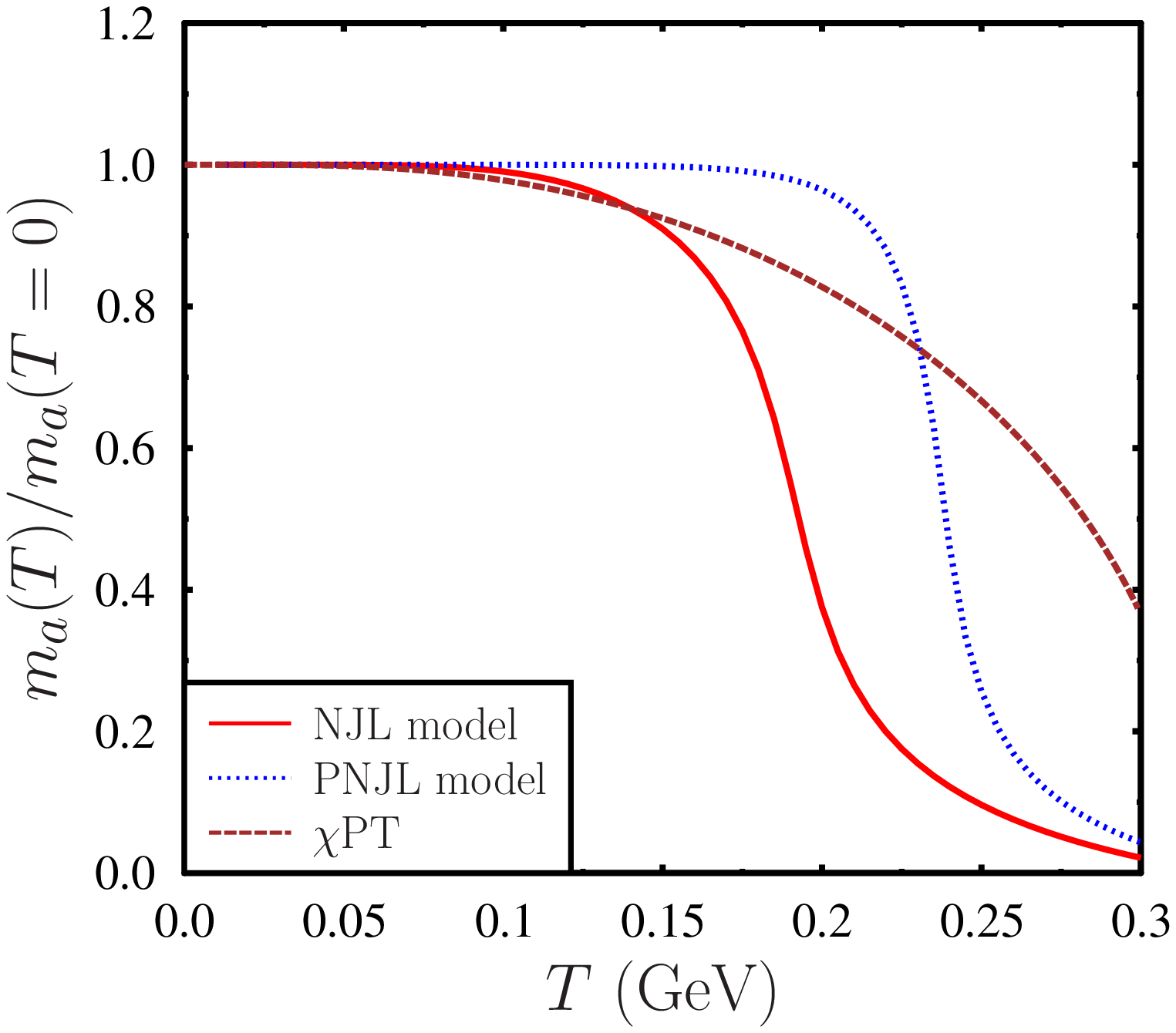}
\end{minipage}%
\begin{minipage}{.5\textwidth}
  \centering
  \includegraphics[width=1.0\linewidth]{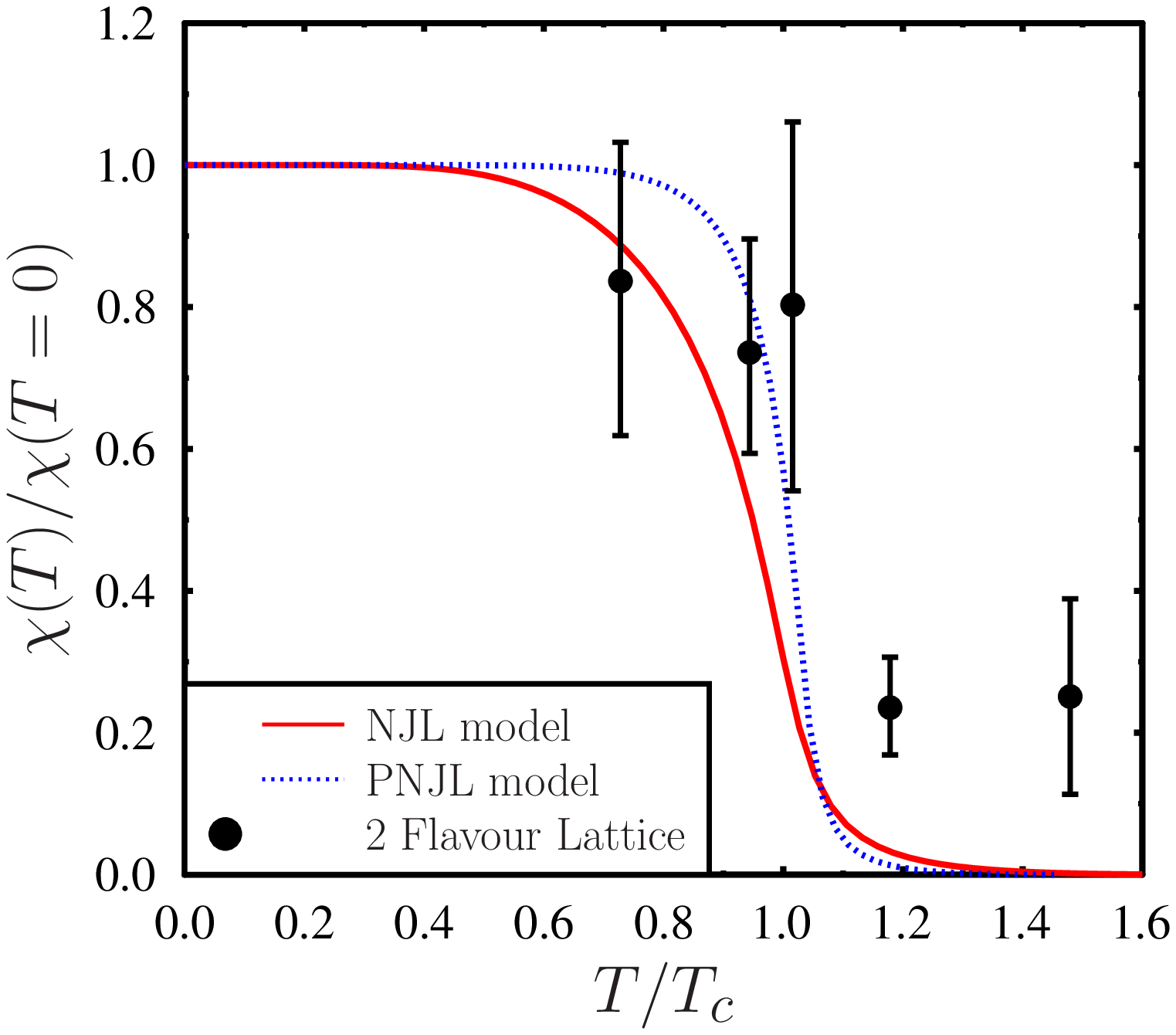}
\end{minipage}
\caption{Left plot: Variation of the normalized axion mass $m_a(T)/m_a(T=0)$ with temperature for PNJL model. $m_a(T=0)$ is the axion mass in vacuum. For comparison we have also plotted the NJL model results and chiral perturbation theory results for the axion mass. Right plot: Variation of the Topological susceptibility $\chi(T)/\chi(T=0)$ with temperature. For comparison we have also provided the results for NJL model and 2 flavour lattice QCD results as given in Ref.\cite{Lattice}.}
\label{fig8}
\end{figure}

Next we show the variation of the normalized or dimensionless axion mass ($m_a(T)/m_a(T=0)$) and the normalized topological susceptibility ($\chi(T)/\chi(T=0)$) with temperature. In the left plot of Fig.\eqref{fig8} we show the variation of the normalized axion mass ($m_a(T)/m_a(T=0)$) for PNJL model. For a comparison, we have given the normalized axion mass which is found in the NJL model and in chiral perturbation theory\cite{NJLaxion1, NJLaxion2}. In the next-to-leading order axion mass can be expressed as \cite{chipt},
\begin{align}
\frac{m_a^2(T)}{m_a^2}= 1- \frac{3}{2}\frac{T^2}{f_{\pi}^2}J_1\left(\frac{m_{\pi}^2}{T^2}\right),
\label{equ20}
\end{align}
here, 
\begin{align}
 J_n\left(\zeta\right) = \frac{1}{(n-1)!}\left(-\frac{\partial}{\partial\zeta}\right)^n J_0\left(\zeta\right), ~~~J_0\left(\zeta\right) = -\frac{1}{\pi^2}\int_0^{\infty} dq~q^2 \log\left(1-\exp\left(-\sqrt{q^2+\zeta}\right)\right)
 \label{equ21}
\end{align}
Using Eq.\eqref{equ20} and Eq..\eqref{equ21} the axion mass at finite temperature can be expressed as, 
\begin{align}
 \frac{m_a^2(T)}{m_a^2} = 1-\frac{3}{4\pi^2}\frac{T^2}{f_{\pi}^2}\int_0^{\infty} dq~\frac{q^2}{\sqrt{q^2+\frac{m_{\pi}^2}{T^2}}}\frac{1}{\exp\left({\sqrt{q^2+\frac{m_{\pi}^2}{T^2}}}\right)-1}.
 \label{equ22}
\end{align}
Here $m_a$ is the axion mass in vacuum.
It is important to note that the axion mass as given in the Eq.\eqref{equ22} in the chiral perturbation theory in only applicable in the temperature range below the QCD transition scale.
In the low-temperature range $T\lesssim 100$ MeV there is not much difference between the axion mass for NJL model, PNJL model and chiral perturbation theory. In fact, at zero temperature the axion mass as obtained in the PNJL model is $m_af_a = 0.00638$ GeV$^2$, which is also the value of the axion mass obtained in the NJL model \cite{NJLaxion2}. Note that the value of the axion mass at zero temperature as found in the NJL and PNJL model is similar to the other estimates of the axion mass, e.g. in Chiral perturbation theory \cite{chipt}.  Only in the high-temperature range ($T>100$ MeV) there is a significant difference between the results between NJL and PNJL model. This is because in the high-temperature range the Polyakov loop plays an important role in the PNJL model, where the scalar and the pseudoscalar condensate are affected by the $\Phi$. From this figure, it is clear that the axion mass is sensitive to the chiral transition temperature. In the PNJL model the Polyakov loop affects the chiral transition temperature. In fact chiral transition temperature increases in the PNJL model compared to the NJL model. This increase in the chiral transition temperature is also manifested in the variation of the axion mass with temperature.  Note that according to Eq.\eqref{equ13}, $\chi(T)/\chi(T=0)=m_a^2(T)/m_a^2(T=0)$. Hence the topological susceptibility is just another way to represent the axion mass. In the right plot of Fig.\eqref{fig8} we present the variation of the normalized topological susceptibility $\chi(T)/\chi(T=0)$ with $T/T_c$ in the PNJL model. For comparison, we have also given the results for the NJL model and 2 flavour lattice QCD results \cite{Lattice}. Near the transition temperature PNJL model results for the normalized topological susceptibility is consistent with the lattice QCD results.

\begin{figure}
    %\centering
    \includegraphics[width=0.7\textwidth]{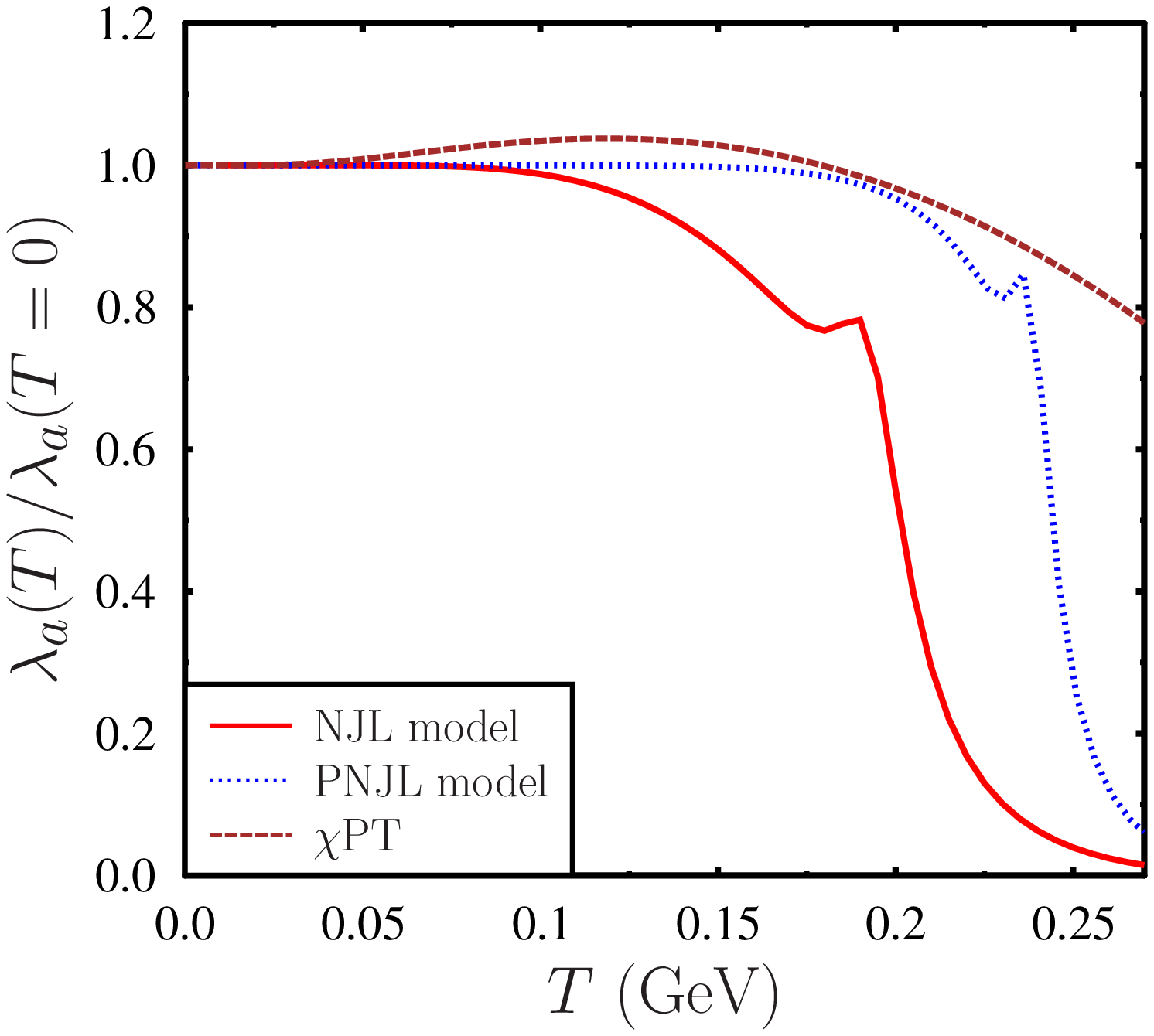}
    \caption{Variation of the normalized axion self coupling $\lambda_a(T)/\lambda_a(T=0)$ with temperature for PNJL model. $\lambda_a(T=0)$ is the axion self coupling in vacuum. For comparison we have also plotted the NJL model results and the chiral perturbation theory results for the axion self coupling.}
    \label{Fig.10}
\end{figure}

Variation of the normalized axion self coupling $\lambda_a(T)/\lambda_a(T=0)$ with temperature is shown in Fig.\eqref{Fig.10}. For comparison, we have also given the results for $\lambda_a(T)/\lambda_a(T=0)$ estimated in the NJL model and in the chiral perturbation theory\cite{NJLaxion1, NJLaxion2}. In the low temperature range as compared to the QCD transition scale the temperature dependent axion self coupling can be expressed as \cite{chipt},
\begin{align}
 \frac{\lambda_a(T)}{\lambda_a} = 1-\frac{3}{2}\frac{T^2}{f_{\pi}^2}J_1\left(\frac{m_{\pi}^2}{T^2}\right)+\frac{9}{2}\frac{m_{\pi}^2}{f_{\pi}^2}\frac{m_um_d}{m_u^2-m_um_d+m_d^2}J_2\left(\frac{m_{\pi}^2}{T^2}\right).
 \label{equ23}
\end{align}
$J_1\left(\frac{m_{\pi}^2}{T^2}\right)$ and  $J_2\left(\frac{m_{\pi}^2}{T^2}\right)$ in Eq.\eqref{equ23} can be obtained using Eq.\eqref{equ21}. Using Eq.\eqref{equ21} and Eq.\eqref{equ23} the axion self coupling at finite temperature for $m_u=m_d$ can be expressed as, 
\begin{align}
\frac{\lambda_a(T)}{\lambda_a} =& 1-\frac{3}{4\pi^2}\frac{T^2}{f_{\pi}^2}\int_0^{\infty} dq~\frac{q^2}{\sqrt{q^2+\frac{m_{\pi}^2}{T^2}}}\frac{1}{\exp\left({\sqrt{q^2+\frac{m_{\pi}^2}{T^2}}}\right)-1}\nonumber\\
& ~~~~+ \frac{9m_{\pi}^2}{8\pi^2f_{\pi}^2}\int_0^{\infty} dq ~\frac{q^2}{\left({\sqrt{q^2+\frac{m_{\pi}^2}{T^2}}}\right)^3}\left( \frac{1}{\exp\left({\sqrt{q^2+\frac{m_{\pi}^2}{T^2}}}\right)-1}\right)^2\nonumber\\
& ~~~~~~~~~~~~~~~~~~~\times\bigg[\exp\left({\sqrt{q^2+\frac{m_{\pi}^2}{T^2}}}\right)\left(\sqrt{q^2+\frac{m_{\pi}^2}{T^2}}+1\right)-1\bigg]
\end{align}
At zero temperature the value of the axion self-coupling as estimated in the PNJL model is, $\lambda_af_a^4= -(55.64)^4$ MeV$^4$. It is clear from the Fig.\eqref{Fig.10} that the results for the NJL model, chiral perturbation theory and PNJL model are in agreement in the low temperature range. But for the high-temperature range, the PNJL model results are significantly different from that of NJL model results.
As mentioned earlier this is due to the fact that in the PNJL model the Polyakov loop $\Phi$ only becomes effective in the high-temperature range $T>100$ MeV.

\subsection{Finite quark chemical potential }
In this subsection, we discuss the results for nonvanishing values of quark chemical potential ($\mu$). It is important to mention that effects of finite
baryon density has been recently discussed in Ref.\cite{axionfinitemu}. At very high densities accessible inside the core of neutron stars the Fermi momentum is not far from the scale of QCD dynamics ($\Lambda_{QCD}$). In particular for the chemical potential above the pion mass, finite density corrections can have an important impact on the properties of the axion \cite{axionfinitemu}. It has been argued in Ref.\cite{axionfinitemu} in the context of chiral Lagrangian extrapolated at finite density,  that due to the suppression of the QCD-instantons the axion mass decreases in a dense medium with respect to the vacuum (without medium). We also get a suppression in the axion mass and coupling as estimated at finite quark chemical potential. 

\begin{figure}
\centering
\begin{minipage}{.5\textwidth}
  \centering
  \includegraphics[width=1.05\linewidth]{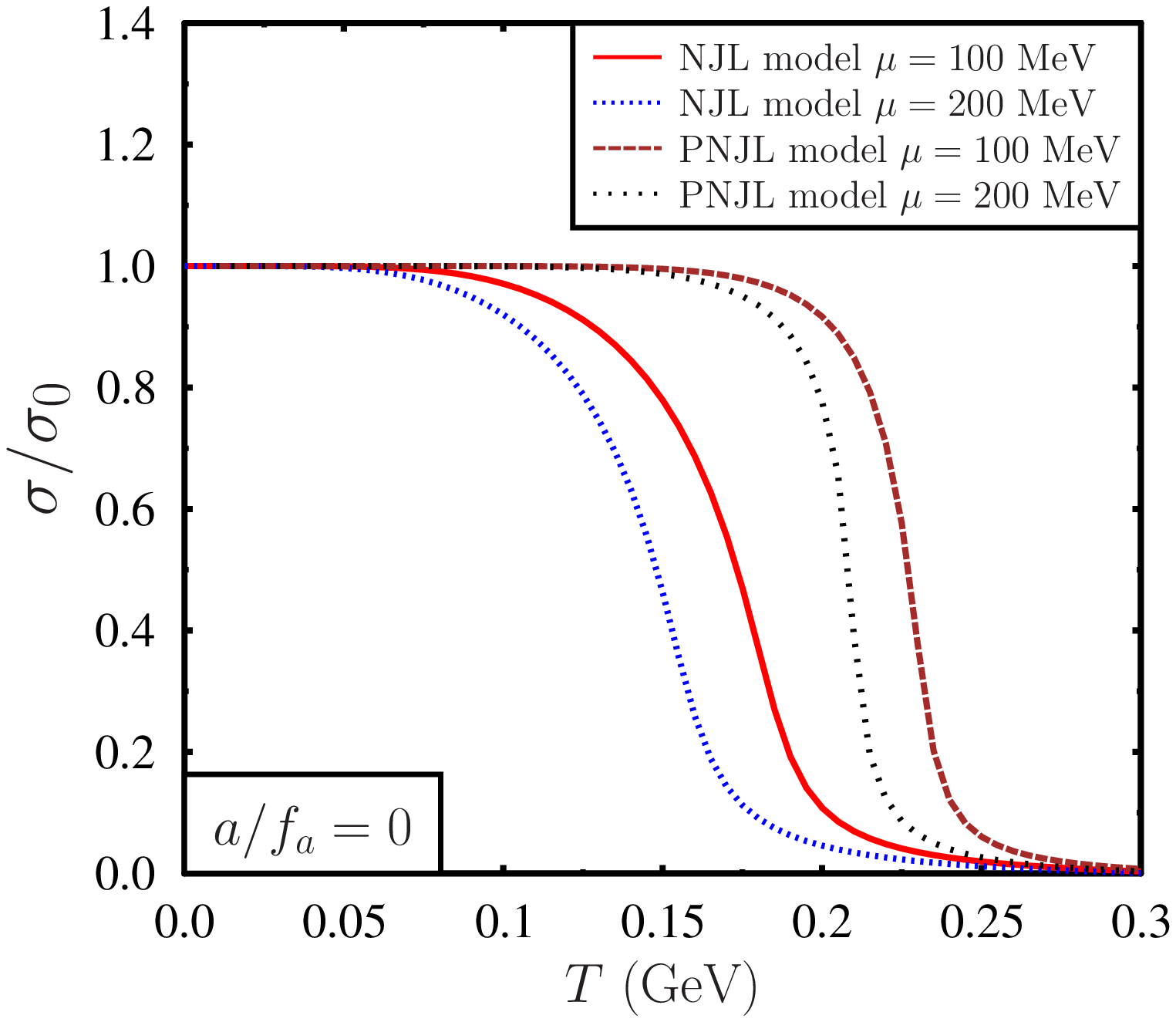}
\end{minipage}%
\begin{minipage}{.5\textwidth}
  \centering
  \includegraphics[width=1.05\linewidth]{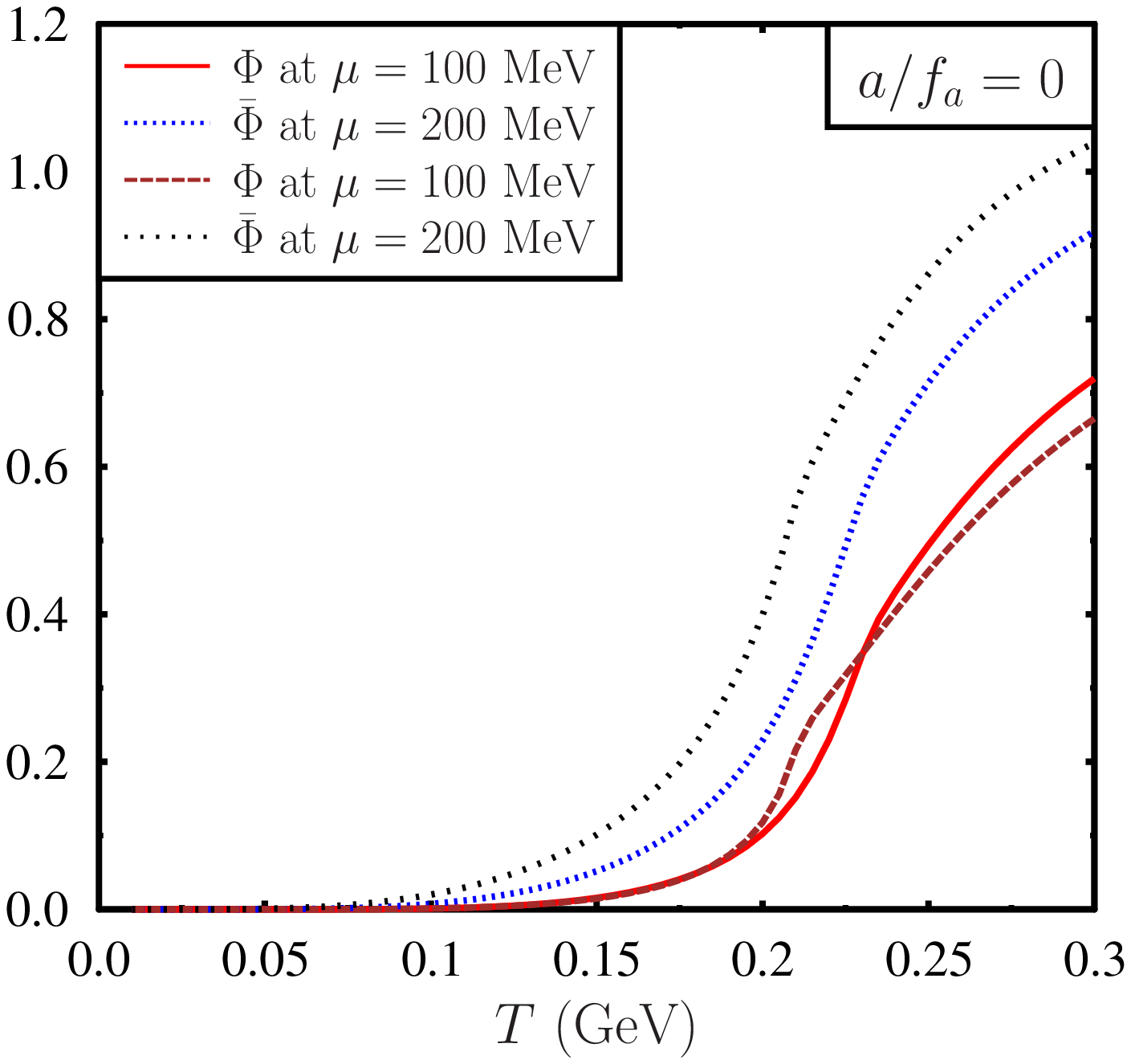}
\end{minipage}
\caption{Left plot: variation of the normalized chiral condensate $(\sigma/\sigma_0)$ with temperature for nonvanishing values of the quark chemical potential. Right plot: variation of the Polyakov loop $\Phi$, and its conjugate $\bar{\Phi}$ with temperature for nonvanishing values of the quark chemical potential.}
\label{Fig11}
\end{figure}

In the left plot in Fig.\eqref{Fig11} we have shown the variation of the normalized chiral condensate $(\sigma/\sigma_0)$. On the other hand the right plot in Fig.\eqref{Fig11} shows the variation of the Polyakov loop $\Phi$, and its conjugate $\bar{\Phi}$ with temperature for nonvanishing values of the quark chemical potential. Note that for Fig.\eqref{Fig11} we have considered the value of $a/f_a =0$, this is because here we are interested in the axion properties which are defined at $a/f_a=0$. The left plot in Fig.\eqref{Fig11} indicates that with an increase in the quark chemical potential the chiral transition temperature decreases both in NJL and PNJL model. Note that for the range of quark chemical potential considered in these plots the chiral transition is smooth. Only at higher quark chemical potential and low-temperature, one expects the chiral transition to be discontinuous \cite{BuballaReview}. From the right plot in Fig.\eqref{Fig11} we see that variation of the $\Phi$ and $\bar{\Phi}$ with the temperature at finite quark chemical potential is similar to the variation of   $\Phi$ and $\bar{\Phi}$ at vanishing quark chemical potential. The only important difference here is the fact that at finite quark chemical potential the values of Polyakov loop $\Phi$ and it's conjugate $\bar{\Phi}$ are not the same. Note that at finite quark chemical potential the thermodynamic potential contains an imaginary part, and the condensates $\Phi$ and $\bar{\Phi}$ are real but not the same. This is reminiscent of the complex fermion determinant for nonvanishing value of the quark chemical potential.

\begin{figure}
\centering
\begin{minipage}{.5\textwidth}
  \centering
  \includegraphics[width=1.0\linewidth]{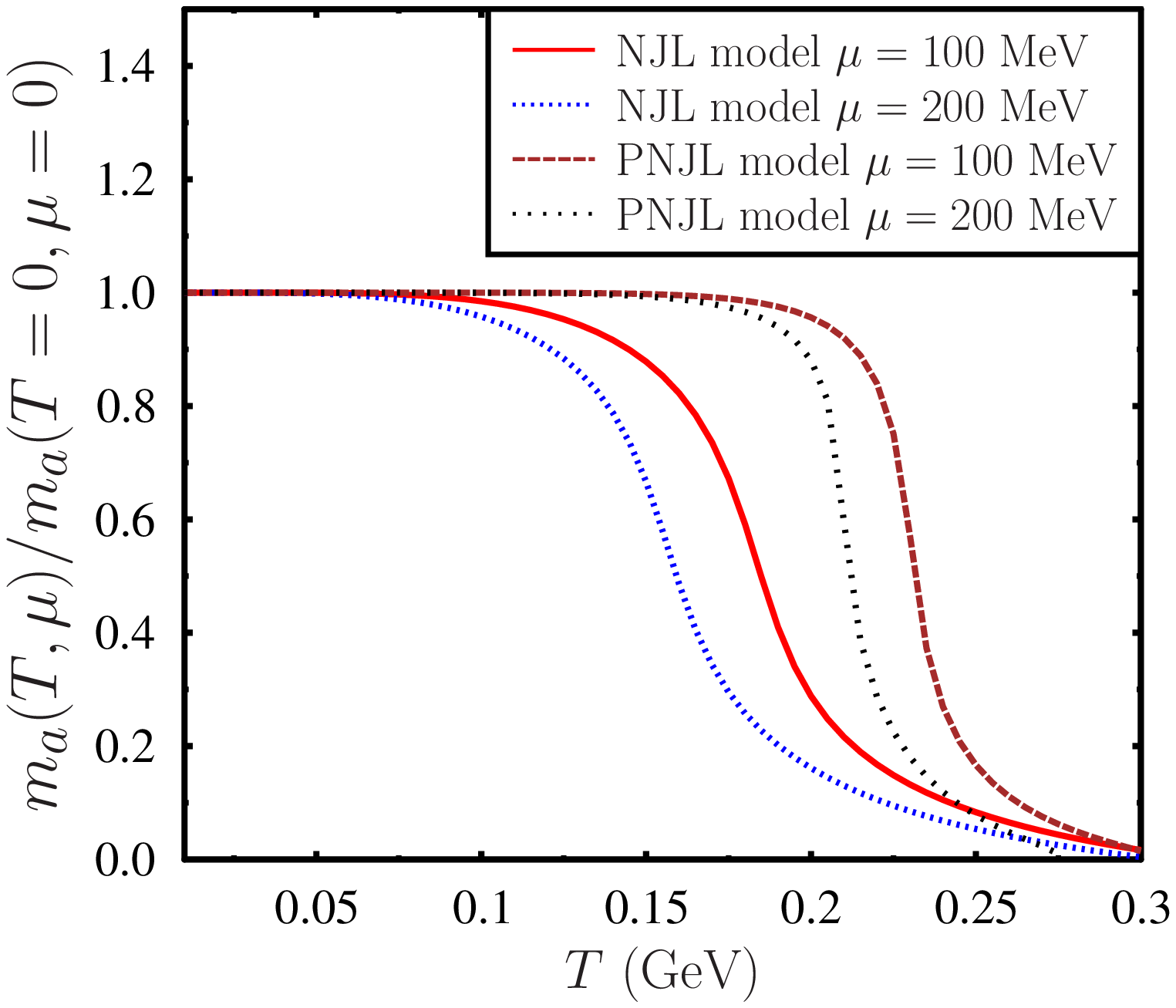}
\end{minipage}%
\begin{minipage}{.5\textwidth}
  \centering
  \includegraphics[width=1.0\linewidth]{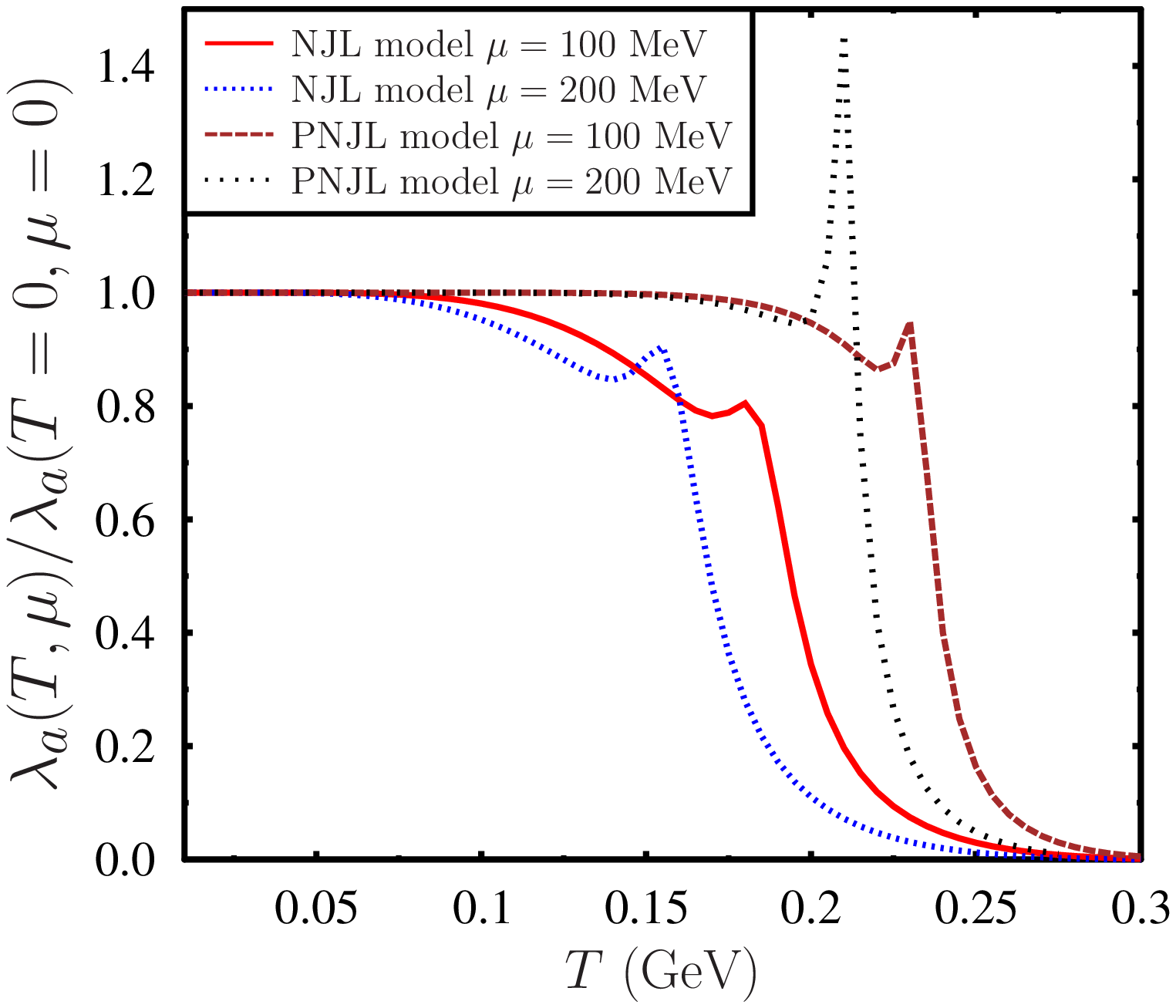}
\end{minipage}
\caption{Left plot: Variation of the normalized axion mass at finite temperature and quark chemical potential with temperature for different values of the quark chemical
 potential. At finite quark chemical potential due to the suppression of the instanton effects the axion mass reduces with quark chemical potential. Right plot: Variation of the normalized axion self coupling at finite temperature and quark chemical potential, with temperature for different values of the quark chemical
 potential. Similar to the axion mass, axion self coupling also reduces with temperature and quark chemical potential.}
\label{fig12}
\end{figure}

Next, we show the variation of the axion properties with temperature for nonvanishing values of the quark chemical potential. In the left plot of Fig.\eqref{fig12} we have shown the variation of the 
normalized axion mass with temperature and quark chemical potential. It is clear that there is a correlation between the scalar condensate and the axion properties, i.e. in the chirally restored or deconfined phase the value of the scalar condensate, as well as the axion mass, is small with respect to the same in the confined or the chiral symmetry broken phase. With increasing quark chemical potential chiral transition temperature decreases which affect the axion mass. Hence with an increase in the quark chemical potential axion mass starts to decrease at a lower temperature both in the NJL and PNJL model. In the right plot, we show the variation of the normalized axion self-coupling, i.e. $\lambda_a(T,\mu)/\lambda_a(T=0,\mu=0)$ with temperature and quark chemical potential. With an increase in the quark chemical potential, it is clear that $\lambda_a(T,\mu)/\lambda_a(T=0,\mu=0)$ starts to decrease at a lower temperature both in NJL and PNJL model. This behaviour of $\lambda_a(T,\mu)/\lambda_a(T=0,\mu=0)$ is consistent with the fact that with an increase in the quark chemical potential chiral transition temperature decreases.   

\begin{figure}[h]
    %\centering
    \includegraphics[width=0.7\textwidth]{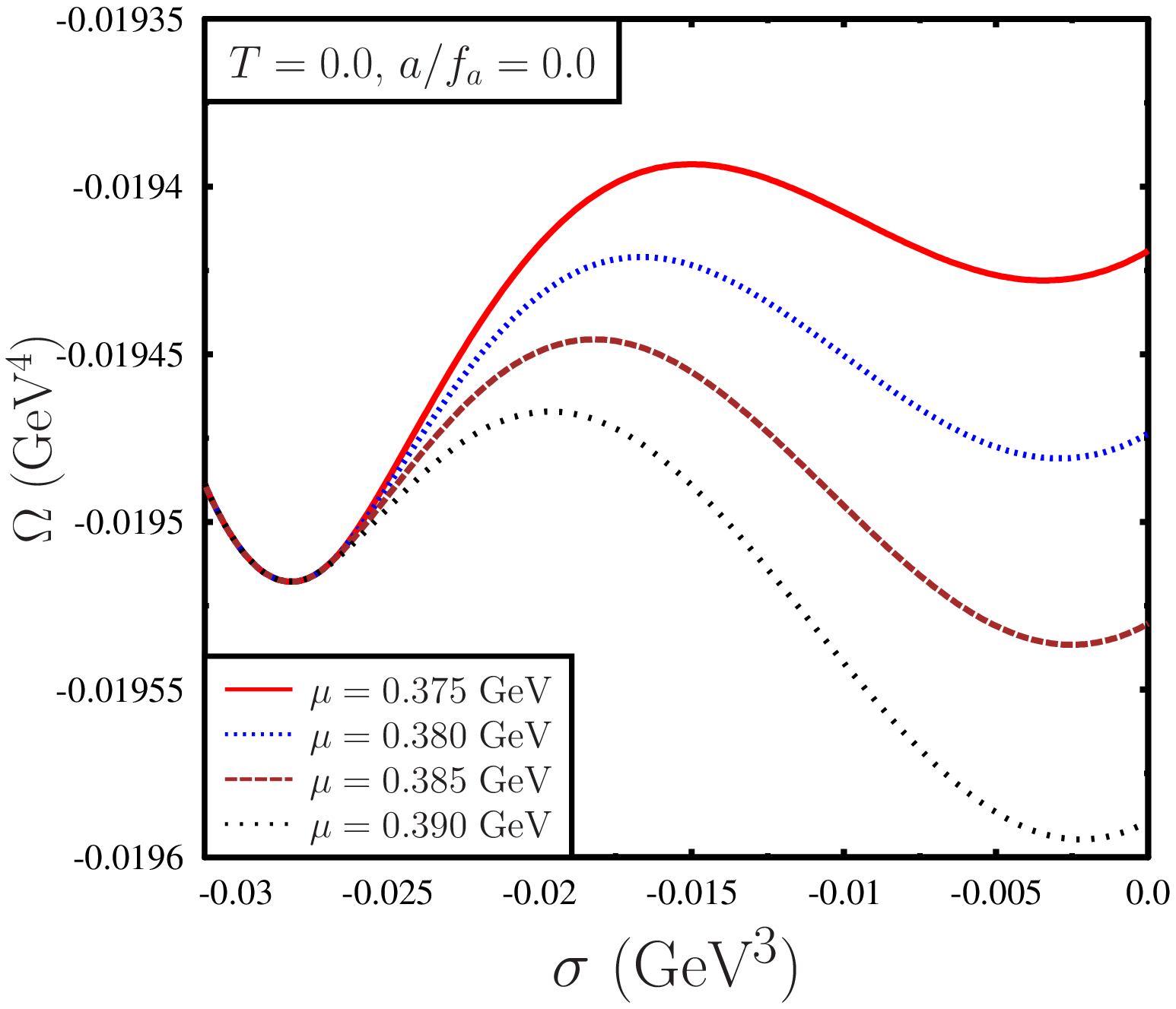}
    \caption{Variation of the thermodynamic potential $\Omega$ with chiral condensate $\sigma$ at $T=0$ for different values of the quark chemical potential. Multiple vacuum structure of the thermodynamic potential is clearly observed in this figure. For $\mu\lesssim 0.385$ GeV the vacuum at $\sigma\sim -0.028$ GeV$^3$ is stable as compared to the vacuum at $\sigma\sim -0.001$ GeV$^3$. But for $\mu> 0.385$ GeV the vacuum at $\sigma\sim -0.001$ GeV$^3$ is stable. It is also clear that across the critical chemical potential the chiral condensate changes discontinuously. At zero temperature the chiral transition is first order in nature and the critical value of this transition is $\mu\sim 0.385$ GeV for the parameters considered here.}
    \label{fig13}
\end{figure}

Finally, we discuss the results for vanishing temperature and finite quark chemical potential. Note that for zero temperature and $a/f_a=0$ all the condensate apart from the scalar condensate, i.e. $\eta$, $\Phi$ and $\bar{\Phi}$ does not contribute to the thermodynamic potential. Hence for zero temperature, there is no difference between the NJL and PNJL results.  In Fig.\eqref{fig13} we have shown the variation of the thermodynamic potential with the scalar condensate ($\sigma$) for different values of quark chemical potential. From this figure, we can see that the thermodynamic potential has multiple vacuum structure. At $\mu \lesssim 0.385$ GeV the vacuum at $\sigma\sim-0.028$ GeV$^3$ is stable and the vacuum at $\sigma\sim-0.001$ GeV$^3$ is not a stable vacuum.
However, at $\mu\sim 0.385$ GeV both the vacuum are almost degenerate. The presence of degenerate vacua in the thermodynamic potential is the indication of the first-order phase transition. For $\mu>0.385$ GeV the vacuum near $\sigma\sim-0.001$ GeV$^3$ is a stable vacuum. Therefore at zero temperature and finite quark chemical potential, the chiral transition is first order in nature, because the chiral condensate which is also the order parameter of the chiral transition changes discontinuously across the critical value of the quark chemical potential. In this case, the first order chiral phase transition at zero temperature occurs at  $\mu\sim 0.385$ GeV.

\begin{figure}
\centering
\begin{minipage}{.5\textwidth}
  \centering
  \includegraphics[width=1.1\linewidth]{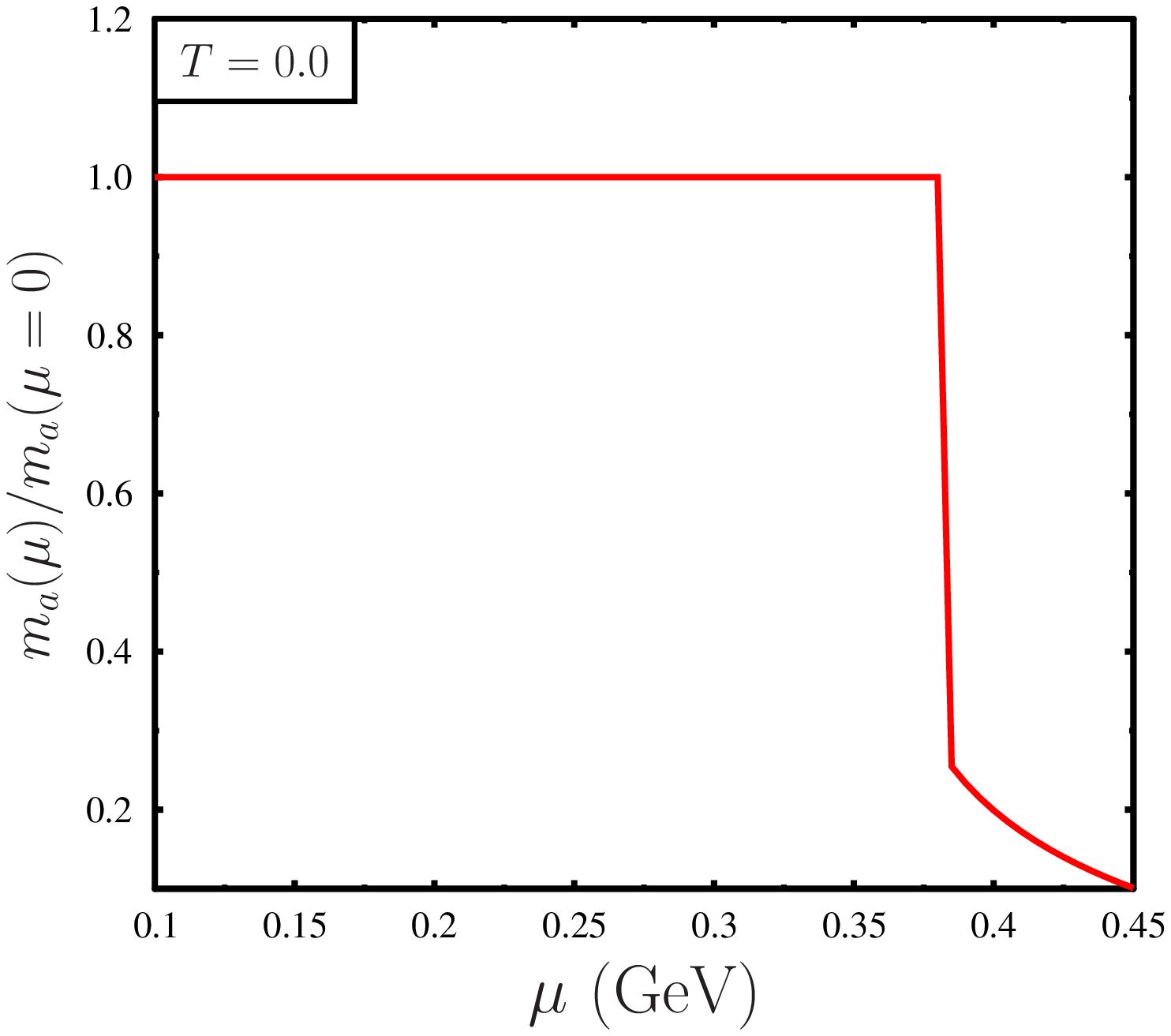}
\end{minipage}%
\begin{minipage}{.5\textwidth}
  \centering
  \includegraphics[width=1.1\linewidth]{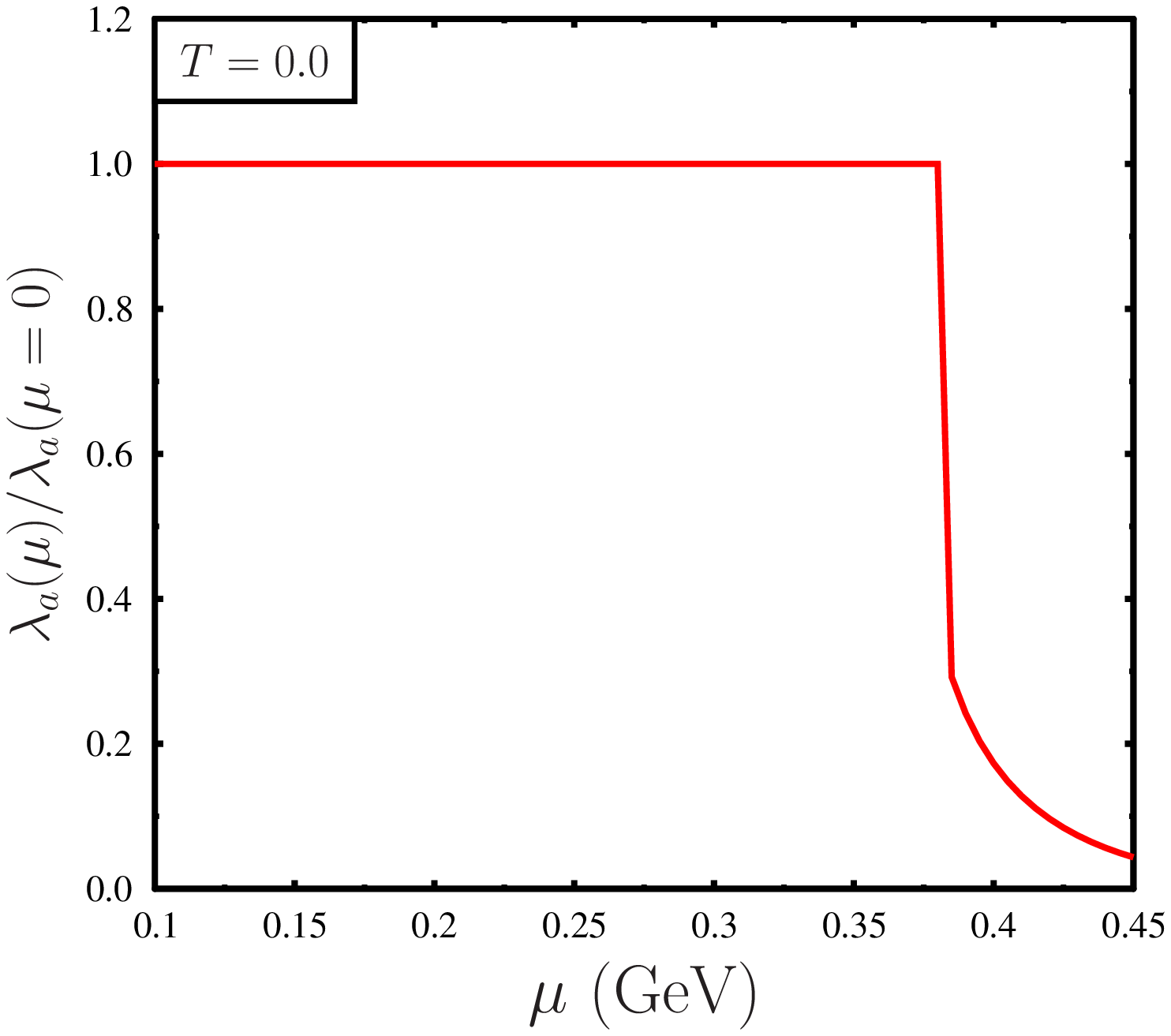}
\end{minipage}
\caption{Left plot: variation of the normalized axion mass $m_a(\mu)/m_a(\mu=0)$ with quark chemical potential ($\mu$) at zero temperature for $a/f_a=0$. Right plot: variation of the normalized self coupling of axion $\lambda_a(\mu)/\lambda_a(\mu=0)$ with quark chemical potential ($\mu$) at zero temperature for $a/f_a=0$. }
\label{fig14}
\end{figure}

The discontinuous change of the chiral condensate also translates into the variation of the normalized mass of the axion ($m_a(\mu)/m_a(\mu=0)$) and the axion self coupling ($\lambda_a(\mu)/\lambda_a(\mu=0)$). In the left plot of Fig.\eqref{fig14} we have shown that variation of the normalized mass of the axion ($m_a(\mu)/m_a(\mu=0)$) the with quark chemical potential
at zero temperature. On the other hand in the right plot of Fig.\eqref{fig14} we have shown that variation of the normalized axion self-coupling ($\lambda_a(\mu)/\lambda_a(\mu=0)$) the with quark chemical potential at zero temperature. Due to the correlation between the chiral transition and the axion properties, we can see a discontinuous variation of the axion properties across the critical value of the quark chemical potential. 

\section{CONCLUSION}
\label{conclusion}

In this article, we have studied the effects of hot and dense QCD medium on the axion properties e.g. it's mass and self coupling. In this investigation QCD medium at finite temperature and finite quark chemical potential has been described by Polyakov loop enhanced Nambu-Jona-Lasinio (PNJL) model. In Ref.\cite{NJLaxion1,NJLaxion2} the 
temperature dependence of axion mass and coupling has been investigated within the framework of Nambu-Jona-Lasinio model. Various methods have been used in literature 
to study the axion properties at finite temperature. For low temperature range chiral perturbation theory ($\chi$PT) gives reliable prediction of the axion mass and self coupling. On the other hand at high temperature as compared to the QCD transition temperature perturbative techniques can be used to estimate the finite temperature effect  on the axion mass and coupling. However near the QCD transition scale where non perturbative techniques are important one can use the QCD inspired effective models to describe the hot and dense QCD medium. NJL model which is one of such QCD inspired models indicates that across the QCD transition temperature the axion properties can be significantly modified by the hot and dense medium. However it is important to note that although the chiral transition or the phenomenological aspects of the chiral symmetry of QCD is incorporated in the NJL model, but due to the lack of the QCD gauge fields, confinement property of QCD is not properly incorporated in the NJL model. Polyakov loop enhanced Nambu-Jona-Lasinio model, on the other hand, takes into account the phenomenological aspects of the chiral symmetry of QCD along with confinement in an unified framework.  
The Polykaov loop which takes a nonvanishing value at finite temperature and quark chemical potential plays an important role near the critical temperature in PNJL model. Therefore there is a significant difference in axion properties calculated in NJL
and PNJL model due to nonzero value of Polyakov loop around the 
critical temperature.  We find that axion properties are correlated with the chiral transition or confinement-deconfinement transition. The axion mass, topological susceptibility and self coupling etc., are significant upto a  temperature $T\sim 230$ MeV in PNJL model at vanishing quark chemical potential as compared to the NJL model where the axion mass and self coupling becomes small at a relatively small temperature $T \sim 190$ MeV. These properties show a weak temperature dependence at low temperature as compared to the QCD transition temperature and they changes significantly across the QCD transition scale.  We have compared the normalized susceptibility ( ratio of 
susceptibility at finite temperature to the susceptibility at zero temperature) at vanishing quark chemical potential with Lattice QCD results. It matches well with 
lattice QCD results. We have also calculated 
axion properties at finite chemical potential. The critical temperature for chiral symmetry restoration decreases as chemical potential increases which also translates into the axion properties, i.e. axion mass as self coupling decreases with chemical potential. At zero temperature but for nonvanishing quark chemical potential, different axion properties like mass and self coupling  are constant up to the critical chemical potential and decreases above the critical chemical potential. At zero temperature but finite quark chemical potential the axion mass and self coupling changes discontinuously across the QCD transition scale. Axions are discussed in literature in various context, e.g. early universe, dark matter, neutron stars etc. It is important to note that in the context of early universe the quark chemical potential or the baryon chemical potential is very small. Therefore in this case the effect of hot QCD medium on the axion properties can be important input for the axion physics. On the other hand inside the compact objects like neutron stars the temperature is rather small but the baryon number density can be large. In this scenario the effect of quark chemical potential on the axion properties can be significant in the study of the axion physics.   

\section*{ACKNOWLEDGMENTS}
The work of A.D. is supported
by the Polish National Science Center Grants No. 2018/30/E/ST2/00432.

\end{document}